\documentclass{article}
\usepackage{arxiv}
\usepackage[utf8]{inputenc} 
\usepackage[T1]{fontenc}    
\usepackage{booktabs}       
\usepackage{amsfonts}       
\usepackage{nicefrac}       
\usepackage{microtype}      
\usepackage{graphicx}
\usepackage{amssymb}
\usepackage{amsmath}
\usepackage{doi}
\usepackage{rotating}
\usepackage{hyperref}       
\hypersetup{
    pdftitle={Calculation of the High-Energy Neutron Flux for Anticipating Errors and Recovery Techniques in Exascale Supercomputer Centres},
    pdfauthor={Hernán Asorey, Rafael Mayo-García},
    pdfkeywords={neutron flux, supercomputing, HPC, exascale, atmospheric radiation},
    colorlinks={true},
    linkcolor={blue},
    citecolor={blue},
    urlcolor={blue}
}
\usepackage{cleveref}       

\title{Calculation of the High-Energy Neutron Flux for Anticipating Errors and Recovery Techniques in Exascale Supercomputer Centres}
\author{
    \href{https://orcid.org/00000-0002-4559-8785}{\includegraphics[scale=0.06]{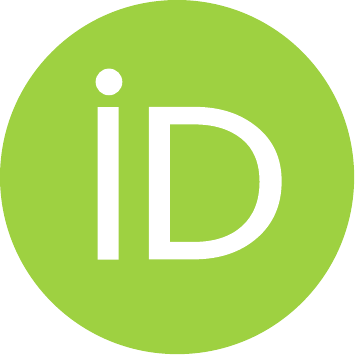}}\hspace{1mm}Hernán~Asorey\thanks{corresponding author:\href{mailto:hernanasorey@cnea.gob.ar}{hernanasorey@cnea.gob.ar}}\\
    Medical Physics Department \& Instituto de\\
    Tecnologías en Detección y Astropartículas\\
    Comisión Nacional de Energía Atómica\\
    Centro Atómico Bariloche\\
    Av. E. Bustillo 9500\\
    8400 San Carlos de Bariloche, Argentina\\
    \\
	\And
	\href{https://orcid.org/0000-0002-0151-3954}{\includegraphics[scale=0.06]{orcid}}\hspace{1mm}Rafael Mayo-García\\
	Technology Department\\
	Centro de Investigaciones Energéticas, \\
    Medioambientales y Tecnológicas (CIEMAT)\\
	Av. Complutense 40\\
    28040 Madrid, Spain\\
}

\renewcommand{\headeright}{}
\renewcommand{\undertitle}{}
\renewcommand{\shorttitle}{High-Energy Neutron Flux and Soft Error Rates at Exascale Centres}
\newcommand{\overbar}[1]{\mkern 1.5mu\overline{\mkern-1.5mu#1\mkern-1.5mu}\mkern 1.5mu}

\begin{document}
\maketitle
\begin{abstract}
    The age of exascale computing has arrived and the risks associated with neutron and other atmospheric radiation are becoming more critical as the computing power increases, hence, the expected Mean Time Between Failures will be reduced because of this radiation. In this work, a new and detailed calculation of the neutron flux for energies above 50~MeV is presented. This has been done by using state-of-the-art Monte Carlo astroparticle techniques and including real atmospheric profiles at each one of the next 23 exascale supercomputing facilities. Atmospheric impact in the flux and seasonal variations were observed and characterised, and the barometric coefficient for high-energy neutrons at each site were obtained. With these coefficients, potential risks of errors associated with the increase in the flux of energetic neutrons, such as the occurrence of single event upsets or transients, and the corresponding failure-in-time rates, can be anticipated just by using the atmospheric pressure before the assignation of resources to critical tasks at each exascale facility. For more clarity, examples about how the rate of failures is affected by the cosmic rays are included, so administrators will better anticipate which more or less restrictive actions could take for overcoming errors. 
\end{abstract}
\keywords{neutron flux \and supercomputing \and HPC \and exascale \and atmospheric radiation}

\section{Introduction}\label{sec:Introduction}

Exascale computing presents several issues, being fault tolerance one of the main ones: while the Mean Time Between Failures (MTBF) of the hardware components (from coolers to memories or random issues) does not grow as fast as the number of resources, the number of cores on a hardware unit experiences continuous growth, and so the probability of one or more tasks being affected by a failure increases\,\cite{dongarra2009international}.
For example, large parallel jobs may fail as frequently as once every 30 minutes on exascale platforms\,\cite{snir2014addressing}.
Also, the higher number of tasks composing a job, the higher will be the computational and economics lost associated with the increasing number in failures.
Although these issues pose enough of a risk, additional factors are now coming into play: clusters with lower energy consumption that are designed and fed with a lower voltage, or smaller circuits are more easily upset because they carry smaller charges and are more prone to hardware failures, or supercomputers (partially) built with GPUs cards counting on an amazing number of cores, or much more complex software being executed, etc.
All the previous results in a higher failure rate, and so, lower values of the MTBF\@.
Thus, there is a necessity in developing tools and frameworks that reduce the impact of tasks and jobs failure on exascale supercomputers.

Traditionally, general fault-tolerant behaviour has been achieved by redundancy and checkpointing mechanisms.
Isolated redundancy is not an ideal approach for HPC as it leads to performance loss, but it has provided nice results in HTC environments (Desktop, Grid, Cloud) or combined with additional methods.
Checkpointing techniques have provided good results on a three-fold basis (system-, user-, and application-level) and have demonstrated a wide scenario of solutions on coordinated and uncoordinated actions, roll-back and roll-forward strategies, mono- and multilevel checkpointing, etc.

Even more and beyond the proper interest of resilience, a consequence of the increase of parallelism both on the hardware and applications sides was a series of problems related to task scheduling.
The idea was to assign tasks to resources trying to avoid starvation, deadlocks, and performance losses, all while having the cluster as full as possible.
This computing efficiency improvement could be achieved by profiting from a proactive (not reactive to failures) checkpointing strategy that could be designed as part of the resource manager scheduler.
For example and among other results, the user-level checkpointing library DMTCP was seamlessly integrated into Slurm\,\cite{rodriguez2019job}.
By designing several dynamic scheduling algorithms and profiting from a new command ({\texttt{smigrate}}), a more resilient system was provided in which also proactive checkpointing actions could be performed for enhancing the computing and energy efficiency by dynamically migrating tasks previously saved with such a checkpoint with low overhead.

This fact has opened the door to new possibilities such as non-invasive maintenance operations, job preemption, more advanced priority policies, lower energy consumption, etc.
Then, further advances must be envisioned once traditional checkpointing and rollback recovery strategies have been accomplished.
In this regard, Silent Data Corruption (SDC) errors, or simply, silent errors (SE) have become a cornerstone in the path to exascale computing.
Soft errors can be mainly classified into two categories: bit-flipping error (e.g.,~1 becomes 0) in RAM; and computation error (e.g.~$1+1=3$) in floating point units.
Traditionally, bit-flipping errors have been handled by the Error Correcting Code (ECC) technique, and computation error is dealt with redundancy methods (ECC cannot handle computation error).
Unlike aforementioned fail-stop failures, such latent errors cannot be detected immediately, and a mechanism to detect and overcome them must be provided as they are becoming a major drawback as the supercomputer complexity grows.
In other words, failures become a normal part of application executions and, among them, SEs are nowadays those with scarce valid solutions properly tested on real environments.

It has been shown that SE are not unusual and must also be accounted for\,\cite{yeom2010strider}.
The cause may be soft efforts in L1 cache, arithmetic errors in the Arithmetic Logic Unit (ALU), (double) bit flips due to cosmic radiation, etc.
The problem is that the detection of a latent error is not immediate, because the error is identified only when the corrupted data is activated.
One must then account for the detection interval required to detect the error in the error recovery protocol.
Indeed, if the last checkpoint saved an already corrupted state, it may not be possible to recover from the error.
Hence, the necessity to keep several checkpoints so a valid one could roll back to the last correct state.
When dealing with SE, however, faults can propagate to other processes and checkpoints, because processes continue to participate and follow the protocol during the interval that separates the occurrence of the error from its detection.

\begin{figure*}[!ht]
    \begin{center}
        \includegraphics[width=0.9\textwidth]{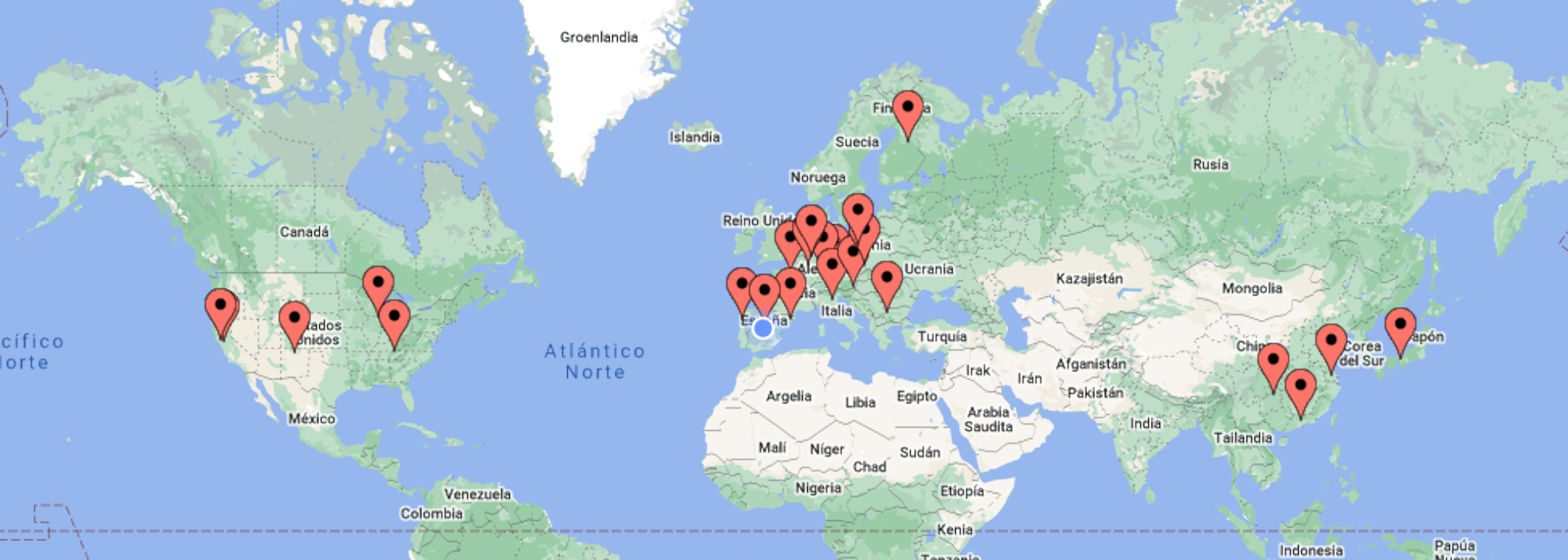}
        \caption{Geographic locations of the 23 exascale supercomputing centres that are being built around the World}\label{fig:map}
    \end{center}
\end{figure*}

Summarizing, there is a clear necessity for overcoming SE as they are becoming inevitable with the ever-increasing system scale and execution time, and new technologies that feature increased transistor density and lower voltage.
Nevertheless, the question of the source for these SE arises.
The answer can be found in the atmospheric cosmic-induced radiation, in which neutrons play a key role.
As neutrons are produced during the interaction of cosmic rays with the atmosphere, and since this last experience seasonal changes, the latitude, longitude, and altitude where a data centre hosts an exascale supercomputer as well as the atmospheric seasonal conditions determine the number of the neutrons reaching the infrastructure and, consequently, the predicted MTBF\@.
So, in this work, using the current techniques for calculating the flux of the expected radiation at the ground originated by the cosmic ray flux, the flux of neutrons with energy $E_n\geq 50$\,MeV averaged per season in 23 data centres are presented.
Among these places, the ones already hosting or expecting to promptly host an exascale supercomputer in China, Europe, Japan, and the United States are included.
The geographic distribution of the 23 exascale supercomputing centres is shown in Figure~\ref{fig:map} and Table~\ref{tab:sites}.

High-energy neutrons, i.e., neutrons with an energy higher than $10$\,MeV, with a total flux of about $13$\,neutrons\,cm$^{-2}$\,h$^{-1}$ in New York at sea level\,\cite{gordon2004measurement,oliveira2017radiation} are expected to cause SE\,\cite{gordon2004measurement}, but the flux of neutrons varies with the geographical location\,\cite{rojdev2015comparison}, altitude\,\cite{dilillo2009neutron}, atmospheric\,\cite{grisales2022impact} and geomagnetic and heliospheric conditions\,\cite{infantino2016fluka}.
As it will be shown later in this work (see the 8$^\mathrm{th}$ column of the table~\ref{tab:sites} in section~\ref{subsec:barometric}), depending on the location the averaged flux of neutrons for $E_n>50$\,MeV could vary between $(3.7\pm0.2)$\,cm$^{-2}$\,h$^{-1}$ in Guangzhou, China, at sea level and $(26.4\pm1.1)$\,cm$^{-2}$\,h$^{-1}$ in Los Alamos, USA, at $2,125$\,m above sea level (asl).

The whole integration of the main source of SE (cosmic radiation) jointly with their prediction process according to the geographical place where such radiation occurs (computing infrastructure location) in a specific season of the year is expected to be useful to the administrators of these supercomputers, given a quantitative measure of the changes in the expected flux of neutrons due to changes in the barometric pressure at the ground level.
With all this information, system administrators will be capable of designing and applying different mathematical and software solutions to cope with these SE that will produce more or less overhead.
This work is expected to be a decision-making tool for the exascale supercomputers' administrators as they will be able to determine in advance which mitigation methodologies need to be applied for overcoming SE depending on the forecasted neutron flux in a specific period of the year.

The main result from this exercise will be a higher resilience, better computational efficiency and less energy misuse in exascale supercomputers.

\section{Related work}\label{sec:related-work}

Fault tolerance can be defined as the capability of a certain system to overcome hardware, software or communication problems and continue with the execution of applications.
This field embraces different sections: the detection of failures, their avoidance if possible, and the recovery from them if not.

To achieve computational resilience, there are several methodologies for overcoming errors produced in runtime.
Technical progress in resilience has been achieved in the last decade, but the problem is not actually solved and the community is still facing the challenge of ensuring that exascale applications complete and generate correct results while running on unstable systems\,\cite{cappello2014toward}.
In this regard, it should be pinpointed that current systems do not have a fully integrated approach to fault tolerance: the different subsystems (hardware, parallel environment software, parallel file system) have their mechanisms for error detection, notification, recovery, and logging.

The current status can be mostly described in a few articles.
In\,\cite{gainaru2013failure}, different approaches towards failure detection and prediction are presented.
A state-of-the-art description of the approaches to overcome these failures is included in\,\cite{cappello2014toward}, where also a more detailed explanation of checkpoint solutions is presented.
An updated status can be found in the compilation of fault detection, fault prediction, and recovery techniques in HPC systems, from electronics to system level, which also analyzes their strengths and limitations and identifies promising paths to meet the reliability levels of exascale systems\,\cite{canal2020predictive}.
These references clearly show that the problem being faced is of real interest in the next generations of supercomputers.

After a failure has been detected (even pre-emptively), checkpoints are a widely used tool devoted to saving the status of the running tasks.
A recent survey of checkpointing protocols can be found in the book edited by Hérault and Robert\,\cite{herault2015fault}.
Strategies also range from coordinated checkpointing (including full and incremental ones) to uncoordinated checkpoint and recovery with message logging, each with different strengths and drawbacks\,\cite{artho2015using}.
Checkpoint pursues to reduce the overhead produced by replication methodologies even when the latter is producing valid results still\,\cite{samfass2021doubt}.

The coordinated checkpoint technique guarantees consistent global states by enforcing each of the processes to synchronize their checkpoints as it is the most common practical choice due to the simplicity of recovery\,\cite{rodriguez2010cppc}.
The obvious issue is to find a balance between the robustness of iterated checkpoints and the induced overhead.
Uncoordinated checkpointing allows different processes to do checkpoints when it is most convenient but is subject to the domino effect, and does not guarantee progress.
Although this issue can be avoided with message logging\,\cite{bosilca2014unified}, uncoordinated checkpointing does not represent a valid alternative in the majority of current production environments and applications.

Recent advances include multi-level approaches, or the use of SSD or NVRAM as secondary storage\,\cite{cappello2014toward} as well as the replication for redundant MPI processes\,\cite{ferreira2011evaluating} and threads\,\cite{yu2011thread}.
Also, on MPI, it is remarkable the initial FT-MPI introduced to enable MPI based software to recover from process failure\,\cite{fagg2000ft} and, also, the enlarged capacities via the Checkpoint-on-Failure protocol for forwarding recovery MPI without resulting in a major overhead\,\cite{bland2013extending}.
Recently, the User Level Failure Mitigation (ULFM) interface provides new opportunities in this field, enabling the implementation of resilient MPI applications, system runtimes, and programming language constructs able to detect and react to failures without aborting their execution\,\cite{losada2020fault}.
Another development is MANA (MPI-Agnostic Network-Agnostic transparent checkpointing) for MPI\,\cite{garg2019mana}, which proposes a new solution especially deserved for exascale\,\cite{xu2021mana}.
The three major approaches to implementing checkpoint systems are application-, user- and system-level (or kernel-level) implementations\,\cite{egwutuoha2013survey}, being the last one always transparent to the user.
The most popular approach is the application-level checkpoint\,\cite{cappello2014toward}, where the programmer defines which is the state to be stored in the application by directly injecting the checkpointing routines directly into the code, or by using some automated pre-processors.
This approach keeps being of interest as new solutions are proposed, such as the application-based focused recovery (ABFR)\,\cite{fang2018abfr}.
This alternative has however been mostly abandoned in the place of the other two, and up to the authors’ knowledge there are currently no significant projects in the area.

With the user-level approach, a library is used to do the checkpointing and the application programs are linked to the library.
User-level does not require system privileges to operate either special kernel modules or kernel patches.
One of the active projects for transparent user-level checkpoints are DMTCP\,\cite{ansel2009dmtcp} or BLCR\,\cite{hargrove2006berkeley}, which include support for distributed and multi-threaded applications and do not require modifying either the application executable or the kernel.

Concerning the state-of-the-art of research on SE, (parallel) jobs can be interrupted at any time for checkpointing, for a nominal cost $C$.
To deal with fail-stop failures, the execution of divisible-load applications is partitioned into same-size chunks followed by a checkpoint, and there exist well-known formulae by Young \& Daly\,\cite{daly2006higher} to determine the optimal checkpointing period.
To deal with SE, the simplest protocol had been to perform a verification (at a cost $V$) just before taking each checkpoint.
If the verification succeeds, then one can safely store the checkpoint and mark it as valid.
If the verification fails, then an error has struck since the last checkpoint, which is correct having been verified, and one can safely recover (which takes a time $R$) from that checkpoint to resume the execution of the application.
This protocol with verifications zeroes out the risk of fatal errors that would force restarting the execution from scratch, but the key point is to find a pattern that minimizes the expected execution time of the application.
Finding the best trade-off between error-free overhead (what is paid due to the resilience method, when there is no failure during execution) and execution time (when errors strike) is not trivial\,\cite{aupy2017coping}.

Later on, it has been published a work for determining the real computational cost in the technique of combining replication and checkpointing\,\cite{benoit2018coping} for assessing either duplication or triplication, which can be acceptable solutions for specific scenarios (aeronautics, for example, though it also requires manufacturing specific hardware as IBM S/390 in Boeing 777\,\cite{yeh1996triple}).
Though it does not specifically try to cope with SE, this work is of interest as it provides closed-form formulas that give the optimal checkpointing period and optimal process count as a function of the error rate, checkpoint cost, and platform size.
Similar work on predicting an optimal checkpointing period and its relationship with the cluster size has been recently published\,\cite{morinigo2019modelling}.

In addition to software techniques, SE can be coped with mathematical approaches.
The traditional wisdom in computing no longer applies as unorthodox, new algorithmic techniques are emerging linked to the exascale requirements.
Aspects related to communicating avoiding algorithms, mixed single-double precision computations or the inclusion of new kinds of randomised algorithms embedded in deterministic portions of the codes are of major concern in the context of faster and more reliable solvers\,\cite{avron2010blendenpik}.

These new methods are insensitive to the quality of the randomness and produce highly accurate results, besides their simplicity and speed\,\cite{baboulin2014using, howell2018iterative}.
Hence, there is currently a large interest in conducting further research on them\,\cite{dongarra2017extreme,chetverushkin2019numerical}.
Specific recent works applied to GMRES\,\cite{morinigo2021error} or parallel stencil computations\,\cite{cavelan2019algorithm} also demonstrate the interest in this topic.

Last but not least, there are some works on radiating computing hardware.
More than twenty years ago, it has been demonstrated that neutrons originated in cosmic radiation are the dominant source of soft errors in DRAM devices\,\cite{normand1996single}, and cosmic-ray induced soft error rates were measured on 16-Mb DRAM memory chips\,\cite{ziegler1998cosmic}.
Later on, in 2002 and 2003, to prove to the manufacturers that the errors appearing in ASC-Q at Los Alamos National Laboratory were due to cosmic rays, the staff placed one of the servers in a beam of neutrons causing errors to spike\,\cite{michalak2005using}.
The Jaguar supercomputer logged single-bit ECC errors at a rate of $350$\,min$^{-1}$ in 2006 as well as double-bit errors once per day, being the latter detected, but not corrected by ECC technique as previously stated.
Also, BlueGene/L at Lawrence Livermore Nat Lab suffered with radioactive lead in the solder to cause bad data in the L1 cache, a problem that ended in slower computations as L1 had to be bypassed.

The  main  effects  of  radiation on semiconductors are the total ionizing dose (TID), the occurrence of Single Event Effects (SEE), and Displacement Damage (DD).
For high-energy neutrons, both the elastic and inelastic interactions are possible, and scattering producing a displacement of atoms from their position in the lattice site results in defects altering the electronic properties of the crystal and being one of the main mechanisms of device degradation\,\cite{lopresti2020neutron}.
The neutron interacts with atoms creating DD and generating secondary charged ionizing particles: a neutron of energy $E_n=100$\,MeV can produce a cascade of secondary particles including secondary neutrons, protons, ions, photons and $\delta$ electrons with energy above $100$\,eV, extending temporal effects and permanent damage far away from the first interaction site\,\cite{han2017characteristics}.
Detailed simulations show that, while the elastic neutron-$^{28}$Si interaction cross-section decreases from $\sim 1,000$\,mb for $E_n \simeq 8$\,MeV down to $\simeq 450$\,mb at $E_n \simeq 100$\,MeV and remains constant up to $E_n\gtrsim1000$\,MeV, the corresponding inelastic cross-section curve starts at $\sim 800$\,mb for $E_n\simeq 10$\,MeV, peaking at $\gtrsim 1,000$\,mb at $\simeq 80$\,MeV and then it stabilizes at $\simeq 200$\,mb for $E_n \gtrsim 1$\,GeV (see Fig.~3 of \,\cite{han2017characteristics}), where $100\,\text{mb}=0.1\,\text{barn}=10^{-25}$\,cm$^2$ means that about $4.2\%$ of the incident neutrons interacts with the $^{28}$Si. Some typical reactions observed involve different mechanisms with energy thresholds between $2.75$ and $12.99$\,MeV, and producing $\alpha$s, such as $^{28}$Si$(n,\alpha)^{25}$Mg and $^{28}$Si$(n,2\alpha)^{21}$Ne, or neutrons, such as $^{28}$Si$(n,n\alpha)^{24}$Mg, or neutrons and protons, such as $^{28}$Si$(n,np)^{27}$Al\,\cite{baumann2005radiation}.
Similar reactions occur with neutrons and oxygen, increasing the probability of having errors with the incident energy as SiO$_2$ is typically in the proximity to active junction areas\,\cite{baumann2001soft}.
Alia {\textit{et al.}}\,\cite{alia2018single} exposed commercial SRAM devices to different flux of protons (30-200\,MeV) and neutrons (5-300\,MeV) and measured the effective $\sigma_\text{err}$ for both types of SEE: soft errors, also known as single event upsets (SEU) in the literature, and hard (or catastrophic) errors just as the single event latch-up (SEL).
By using fitting their experimental data to Weibull functions they compared the $\sigma_{\text{SEU}}$ for neutrons at different energies with the same magnitude for energetic proton, and observed that the behaviour of $\sigma_{n,\text{SEU}}$ depends both on the neutron energy and on the internal geometry of the device, and that $\sigma_{n,\text{SEU}}$ tends to $\sigma_{p,\text{SEU}}$ of protons at $E_p=250$\,MeV for $E_n\gtrsim 25$\,MeV (see Figure~3 of\,\cite{alia2018single}).

As the incident neutron energy gets higher, the number of new reactions in the pathway increases, extending the damage and the probability of having errors from a single reaction.
As it will be detailed in section~\ref{sec:results}, it is possible to characterize the radiation-induced errors in computing devices by defining an effective cross-section, $\sigma_{\text{err}}$, a widely used magnitude to directly evaluate the radiation sensitivity of a particular device\,\cite{baumann2005radiation}.
As it is an effective metric, it considers all the possible sources of neutron-induced computing errors, and it is experimentally measured by placing different devices in a neutron beam and calculating the fraction of the observed rate of neutron-induced errors to the injected neutron flux\,\cite{tiwari2015understanding}.
The Los Alamos Neutron Science Center (LANSCE) irradiation facility is one of the neutron sources typically used to measure the number of fatal soft errors, such as the measurement performed in the ASC-Q supercomputer, one of the world’s fastest supercomputers in 2005\,\cite{michalak2005using}, and in the Titan supercomputer, which is composed of more than $18,000$ Kepler GPUs, has a radiation-induced MTBF in the order of dozens of hours\,\cite{tiwari2015understanding}.

Thus, new works on this SE problem produced by radiation have been more recently published focusing on determining the reliability in GPUs\,\cite{oliveira2017radiation} and Xeon Phis also applying high-level fault injection\,\cite{oliveira2017experimental}, where the relative $\sigma_{\text{err}}$ for each device exposed to high-energy neutrons have been obtained.
Further steps forward have been the comparison between high-energy and thermal neutrons effects on the error rates on Commercial Off-The-Shelf (COTS) devices\,\cite{oliveira2020high} by exposing AMD APUs (4 Steamroller CPUs + 1 AMD Raedon R7), Intel XeonPhi processors, Nvidia K20, TitanX and TitanV GPUs and a Zync-7000 FPGA to two beam of neutrons with energies in the range from $1$\,meV to $1$\,GeV in the ChipIR and Rotax neutron beam-lines at the ISIS Neutron and Muons Source\,\cite{oliveira2020high}.
They conclude that while high-energy neutrons are the most important source of SE, for some applications in some computing devices thermal-neutrons can account up to $59\%$ of the total MTBF\@.
In a latter work, an experimental evaluation of the effective cross-section $\sigma_{\text{err}}$ for a high-energy vs thermal neutron to generate an error in the same computing devices is provided as well as an estimation of the thermal neutrons flux modification due to materials heavily present in a supercomputer room\,\cite{oliveira2021thermal}.

These works also quantify and qualify radiation effects on applications’ output correlating the number of corrupted elements with their spatial locality and provide the mean relative error (dataset-wise) to evaluate radiation-induced error magnitude.
Might it not be forgotten, as transistors get smaller, the amount of energy it takes to spontaneously flip a bit get smaller too, i.e., as exascale arrives, the number of bit-flip errors caused by radiation increases.
Also, previous references about radiating computing hardware are associated to either neutron flux originated in a Lab for quantitatively estimating SE rates or demonstrating how cosmic rays actually affect computations, but what about determining the natural flux that is received in any place in the world?
Hence, the evaluation of the contribution of non-thermal neutrons to the error rate of computing devices can be now calculated for the 23 exascale data centres around the World from the work carried out in the previous references and the results provided in this work.

\section{Atmospheric production of energetic neutrons}\label{sec:the-physical-model}

Cosmic rays are high-energy particles and atomic nuclei with energies from a few GeVs up to $\gtrsim 10^{20}$\,eV\,\cite{bluemer2009cosmic}.
After the pioneering works of Rossi and Auger in the 1930's\,\cite{kampert2012extensive}, it is well established that cosmic rays interact with the atmosphere producing cascades of particles via radiative and decay processes, collectively known as Extensive Air Showers (EAS)\,\cite{grieder2010extensive}.
Depending on the energy $E_p$ of the primary cosmic ray, an EAS could have up to $\sim 10^{10}$ particles at the moment of its maximum development.
The detailed analysis of these phenomena is highly complex, as lot of different processes could be involved as more and more particles are produced.
Essentially, the shower starts in the atmosphere at the first interaction point occurring at an atmospheric depth $X_0$ that depends on primary composition and energy, where it interacts with an atomic nucleus present in the air constituents (see for example\,\cite{abreu2012measurement}).
Due to the enormous difference in the energy when compared with the incoming cosmic ray, the target nuclei can be considered at rest.
Since the transference at these energies of transverse momentum is small, all the increasing number of secondaries are moving towards the ground in the approximate direction of the primary.
However, they can be dispersed, and the small transfer of traverse moment during radiative or decay processes produces a slow drift moving the particles away from the shower axis, and finally remain contained in a curved, thin disk known as the shower front, that moves down to the ground in the direction pointed by the initial momentum of the primary particle.
The distribution of secondary particles in the shower front is axially symmetric and the particle density decrease as a power law with the distance $r$ to the shower axis, being well described by the Nishimura-Kamata-Greisen (NKG) lateral distribution function (LDF)\,\cite{greisen1960cosmic}.

Electromagnetic (EM) showers are initiated by photons or electrons, and most of the processes are mediated by QED interactions.
These cascades are mainly ruled by two interaction channels: (i) $e^\pm$ Bremsstrahlung, and (ii) pair production of $e^\pm$.
It is important to notice that both processes are coupled at high energies, as photons produce $e^\pm$ pairs by (ii), which in turn produce high-energy $\gamma$s by (i).
These processes continue producing EM particles that could initiate new EM sub-cascades and more energy is transferred to the EM channel, which in turn produce new EM secondaries with lower energy.
At some point in the cascade evolution during the propagation through the atmosphere, the rate of occurrence of radiative processes begins to decrease as the mean energy as a function of the atmospheric depth $X$, i.e., $\langle E(X) \rangle = E_p / N(X)$, where $N$ is the total number of secondaries in the cascade, drops below the critical energy $E_c$ and the ionization losses start to dominate over the radiative losses.
At this point, the cascade reaches its maximum development, with a total number of particles $N_{\max}\propto E_p$ and occurring at an atmospheric depth $X_{\max} \propto \log(E_p)$.
The cascade continues collectively moving down to the ground through the atmosphere, and once $X_{\max}$ is surpassed, the total number of particles $N(X)$ starts to monotonically decrease due to: (i) the radiative processes are strongly suppressed for $\langle E(X) \rangle < E_c$; and (ii) the atmospheric absorption raises as the air density increases at lower altitudes.

Instead, a hadron-initiated EAS typically produces new hadrons through fragmentation, and mesons through hadronization of the resulting fragments.
Those mesons, typically $\pi^\pm$ and $\pi^0$, have different energy losses in the air and, most importantly, their corresponding lifetime and decay products are very different, having at the end a major impact on how these cascades develop.
Almost all $\pi^0$, with a lifetime of $\tau_{\pi^0}=8.4\times10^{-17}$\,s\,\cite{Zyla2020particle}, decay very close to their production point into two energetic $\gamma$s that initiate new EM showers, transferring more energy into the EM channel.
Instead, charged pions can propagate through the atmosphere down to typical altitudes of $4-6$\,km due to their longer lifetime $\tau_{\pi^\pm}=2.6\times10^{-8}$\,s\,\cite{Zyla2020particle}.
At these altitudes, they start to decay into charged muons $\mu^\pm$ generating the muonic component of the cascade.
As the shower develops, the energy is continuously transferred to the EM and $\mu$ channels due to the decays of neutral and charged mesons.
Close to the ground, $85-90\%$ of $E_p$ is at the EM channel, and the number of particles ratios typically are $10^2:1:10^{-2}$ for the EM, muon and hadronic channels respectively\,\cite{matthews2005heitler}.
This latter is produced by hadronic interactions and so, it remains close to the shower axis as most of the hadrons move in a close direction to the original one, due to the reduced transference of traverse momentum produced by the leading particle effect of hadronic interactions, see e.g.\,\cite{roberts2007aspects, matthews2005heitler,capdevielle1992influence}.
Therefore, the hadronic component is located in a small region located close to the shower axis and is mainly composed of energetic neutrons and protons, with some light nuclei and charged pions, and small traces of other hadrons.
Neutrons are mainly produced by spallation processes of protons on $^{14}$N and other nuclei in the atmosphere\,\cite{silberberg1990spallation, goldhagen2003cosmic}.
As they are the only quasi-stable neutral hadrons present in the cascade\footnote{It is possible to consider neutron as quasi-stable particles since their lifetime is several orders of magnitude larger than the characteristic time of the cascade evolution.} and no ionization or radiative processes affects their propagation in the atmosphere, their evolution is only determined by elastic and quasi-elastic scattering and hadronic interactions.
As explained in section~\ref{sec:related-work}, the energy distribution of atmospheric neutrons at different places exhibit some similarities and the main variations are related to the location and altitude of the observation site\,\cite{infantino2016fluka,gordon2004measurement,vukovic2010measurements,goldhagen2003cosmic}.
Energy losses in the atmosphere produce two typical structures in the neutron energy spectrum: first, a single peak in the number of muons is observed at $E_n \simeq 100$\,MeV, the so-called quasi-elastic peak;
and a complex structure observed in the $0.1 \lesssim E_n 10$\,MeV caused by many resonances cross-sections depending on the target nuclei.
At lower energies, the spectrum follows a typical $E_n^{-1}$ power law distribution with the neutron energy.
The exact energy at which these spectral features appear depends on several factors, such as the altitude above sea level, geomagnetic field conditions and Solar activity, and the water vapour content in the air\,\cite{clem2004new}.
Due to their energy and the way they propagate through the atmosphere, these neutrons arrive at the ground with a considerable and measurable time delay with respect the primary cascade\,\cite{erlykin2008neutron}.

To properly simulate the cascade evolution and take into account all the involved physical processes and the propagation and tracking of up to $\sim 10^{10}$\,secondary particles is a heavily demanding computing task.
To do so, several tools have been developed, but the most extended and validated one is CORSIKA\,\cite{heck1998corsika}, a program for the detailed simulation of extensive air showers initiated by high-energy cosmic ray particles written in FORTRAN and continuously upgraded\,\cite{engel2019towards}.
However, while it incorporates the possibility to select a specific atmospheric model, the values of the components of the local geomagnetic field and the altitude of the observation level, CORSIKA lacks the possibility to change those values in a dynamic way, or, most importantly, it is not possible to calculate in a direct way the secondary particles at the ground produced by the integrated flux of the primary cosmic rays.
These factors are significant for the calculation of the expected background radiation at any particular site around the World and under specific and time evolving atmospheric and geomagnetic conditions.

When calculating the expected flux of secondary particles, the composition of the primary flux, the local atmospheric profile and its variations along the year, or the secular changes and the fast disturbances introduced by the Solar activity in the Earth's magnetic field have to be taken into account as they affect the number of primaries impinging the Earth's atmosphere, the evolution of the EAS in the air and the consequent flux of secondary particles at the ground.

To accomplish these tasks in a semi-autonomous way, the Latin American Giant Observatory (LAGO)\,\cite{sidelnik2017lago} developed ARTI\,\cite{sarmiento2022arti}, a toolkit designed to effortlessly calculate and analyze the total background flux of secondaries and the corresponding detector signals produced by the atmospheric response to the primary flux of galactic cosmic rays (GCR).
ARTI is publicly available at the LAGO GitHub repository\,\cite{asorey2022arti}.

LAGO operates a network of water Cherenkov detectors (WCD) at different sites in Latin America, spanning over different altitudes and geomagnetic rigidity cutoffs\,\cite{sidelnik2015sites}.
The geographic distribution of the LAGO sites, combined with the new electronics for control, atmospheric sensing, and data acquisition, allows the realisation of diverse astrophysics studies at a continental scale\,\cite{asorey2015lago}.
By using ARTI, LAGO is capable to obtain a better characterization of its distributed detection network and determining the sensitivity to the different phenomena studied, such as the measurement of space weather phenomena\,\cite{asorey2015space} or the observation of high-energy transients\,\cite{sarmiento2021latin}.

ARTI is a computational tool that integrates CORSIKA, Magneto-Cosmic and Geant4 with its own designed control and data analysis codes, allowing the calculation of the expected integrated flux of atmospheric radiation in any geographic location under realistic and time-evolving atmospheric and geomagnetic conditions\,\cite{asorey2018preliminary}.
The expected flux at the ground calculated by ARTI has been contrasted and verified with measurements performed at different astroparticles observatories, as most of them take advantage of the atmospheric muon background for the detector calibration\,\cite{asorey2015lago, pierre2020pierre, galindo2017calibration, pena2021characterization, aab2020studies}.
ARTI also has been extensively used for different applications, such as the characterization of new high altitude sites for the observation of steady gamma sources or astrophysical transients, such as the sudden occurrence of a gamma ray burst~\cite{sarmiento2021latin}; or to study the impact of space weather phenomena from ground level by using water Cherenkov detectors~\cite{asorey2015lago,sarmiento2019modeling,rubio2021eosc}; or to calculate the most statistically significant flux of high-energy muons at underground laboratories~\cite{rubio2021eosc,bertolli2022estimacion}; to help in the assessment of active volcanoes risks in Latin America~\cite{pena2022muography,taboada2022meiga,vasquez2020simulated,vesga2021simuated}; and even to contribute to the detection of improvised explosive devise at warfare fields in Colombia~\cite{vasquez2021improvised}.
In particular, we have used ARTI to estimate the expected response of water Cherenkov detectors, commonly used for astroparticles observation, to the atmospheric neutron flux and its relation with the observation of space weather phenomena\,\cite{sidelnik2020simulation}, and for the design of new safeguard neutron detectors for the identification of traffic of fissile materials~\cite{sidelnik2020enhancing,sidelnik2020neutron}, which involves in both cases the calculation of the expected flux of atmospheric neutrons and the corresponding detector responses\,\cite{sidelnik2020simulation, sidelnik2020neutron}.

Added to the intrinsic complexity of tracking all the relevant interactions of up to billions of particles with the atmosphere just for a single EAS, the atmospheric radiation at the ground level is originated by the convolution of the cascade developments of billions of cosmic rays that simultaneously impinge the Earth's atmosphere.
Therefore, to obtain a statistically significant distribution of secondary particles at the ground, the time integration should be long enough to avoid statistical fluctuations\cite{asorey2015lago,sarmiento2022arti}.
For example, a typical calculation of the expected number of secondaries per square metre per day for a high-latitude site involves the computation of $\sim 10^9$ EAS\@.
For this reason, ARTI is prepared for running at high-performance computing (HPC) clusters operating with the SLURM workload manager, and in Docker containers running at virtualized cloud-based environments such as the European Open Scientific Cloud (EOSC) and capable to store and access the produced data catalogues at federated cloud storage servers~\cite{rubio2021novel, rubio2021eosc}.

In this work previous calculations of the expected flux or particles at the ground level are extended, with special emphasis on the neutron flux, as one of the possible sources of silent and non-silent errors as described in previous sections.
For doing this, we selected the minimum possible available value of the kinetic energy cuts for hadrons in CORSIKA, i.e., $E_{h_{\min}}=5 \times 10^{-2}$\,GeV\@, and so, for the case of neutrons, they have not tracked anymore once they reach this energy limit of $E_{n_{\min}}=50$\,MeV, that corresponds to a total energy of $989.6$\,MeV\@.

As can be inferred from the development of the showers described above, the atmosphere has a crucial role in the final distribution of particles at the ground.
Any atmospheric model describes the atmosphere's main parameters (such as the atmospheric density profile) at a given time and position.
So, to account for the atmospheric impacts on the cascades developments, ARTI can use four different types of atmospheric models:
i) the broad MODTRAN atmospheric model~\cite{Kneizys1996modtran}, that assigns a general profile for different areas of the World depending on latitude and season (tropical, subtropical summer and winter, arctic or antarctic summer and winter)~\cite{Kneizys1996modtran};
ii) local atmospheric profiles based on the Linsley's layers model\,\cite{Linsley1976us} for predefined sites;
iii) extract real-time atmospheric profiles from the Global Data Assimilation\footnote{Data assimilation is the adjustment of the parameters of any specific atmospheric model to the real state of the atmosphere as measured by meteorological observations} System (GDAS)\,\cite{noaa2004global} using the Linsley's model;
and iv) calculate and use the typically monthly-averaged atmospheric profiles for a given location\,\cite{grisales2022impact, rubio2021eosc, rubio2021novel}.
As we will show in the next section, by using these functionalities we can model the expected seasonal variation in the flux of secondary particles at the ground level for each one of the 23 exascale data centres shown in Figure~\ref{fig:map}.

Given all the relevant primaries are charged particles and nuclei, another important factor that should be taken into account is the secular variation of the Earth's magnetic field (EMF) and its fast disturbances.
These effects could be significant for the case of high latitude sites, such as the CSC Kajaani data centre in Finland.
As it is described in\,\cite{asorey2018preliminary,sarmiento2022arti}, ARTI incorporates specific modules to calculate the status of the EMF by using the different EMF models taken into account both the secular variation of the EMF and its disturbances.

In the next section, we show the expected flux of atmospheric radiation at the ground and its corresponding seasonal variations for the 23 exascale supercomputing centres.

\section{Results and Discussions}\label{sec:results}

\subsection{Barometric effects in the flux of high-energy neutrons}\label{subsec:barometric}

The first step in the calculation of the expected flux at the ground is to obtain the magnetic field components $B_x$ (north component) and $B_z$ (vertical component) from the current version of the International Geomagnetic Field Reference (IGRF) model (IGRF13-2019)\,\cite{alken2021international}.
To reduce the impact produced by Solar activity, all the calculations were performed using the configuration of the EMF for December, 20th, 2021, as no disturbances in the magnetosphere were observed for this day.

\begin{figure*}[!ht]
    \begin{center}
        \includegraphics[width=0.495\textwidth]{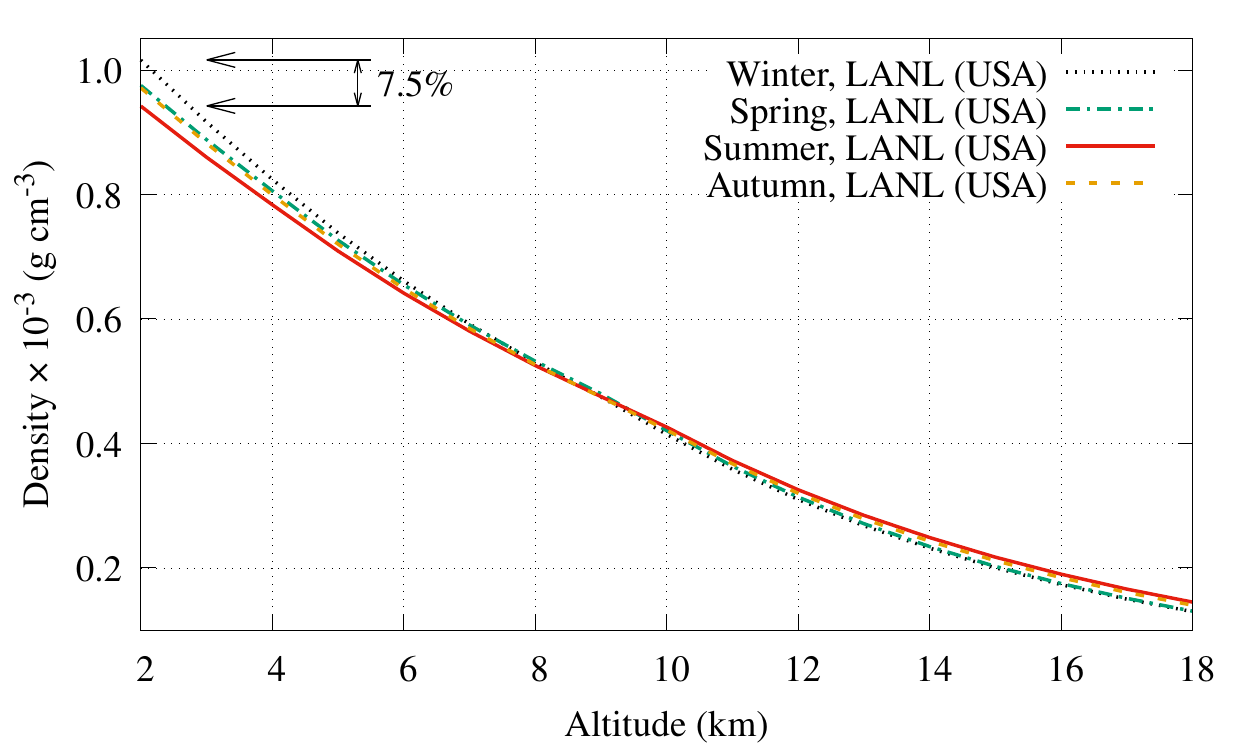}
        \includegraphics[width=0.495\textwidth]{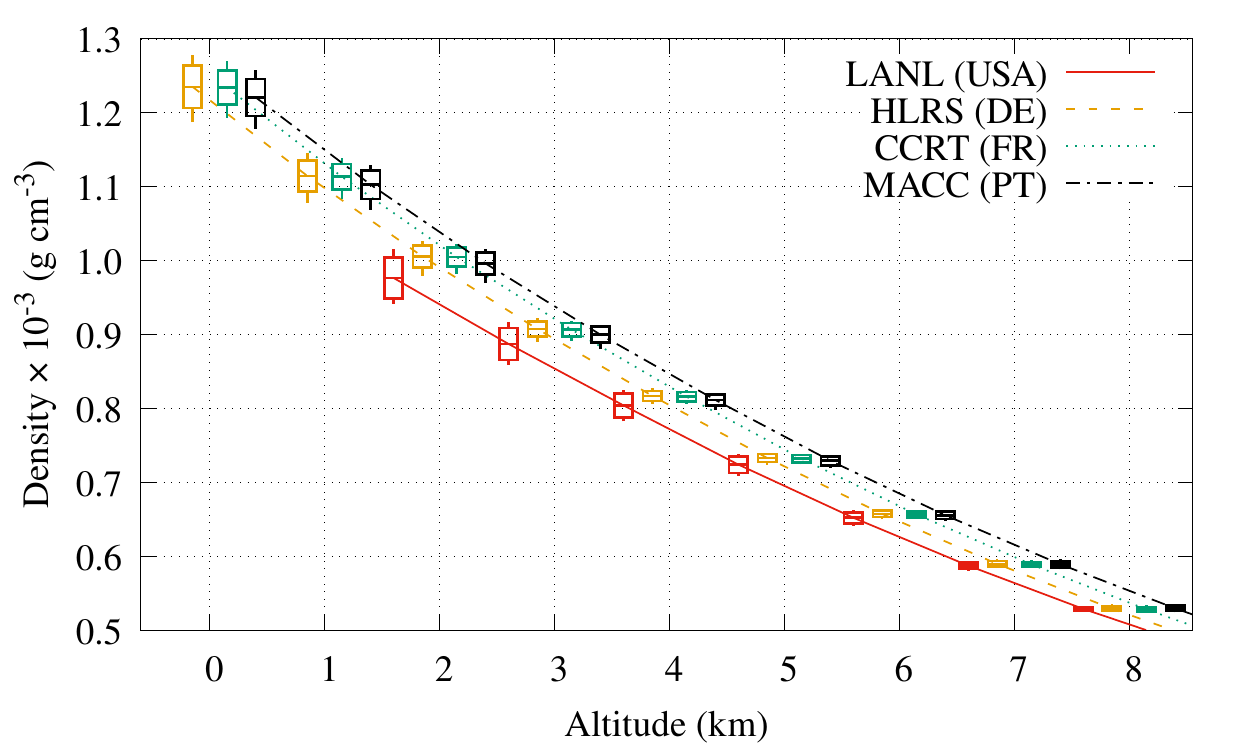}
        \caption{Left: The atmospheric density profiles for LANL are shown for the Winter (dotted black line), Spring (dash-dotted green line), Summer (solid red line) and Autumn (dashed yellow line) of 2020. These profiles were extracted from the GDAS database and averaged for each month. Differences of up to $7.5\%$ can be observed in the density at the ground level in the LANL site, at an altitude of $2,125$\,m asl. Atmospheric profiles used extends up to an altitude of $\sim 110$\,km, corresponding to the limit of the Earth's atmosphere according to Linsley's atmospheric model\,\cite{Linsley1976us}.\\
        Right: Density variations observed at different altitudes along 2020 at LANL (solid red line), HLRS (dashed yellow line), CCRT (dotted green line) and MACC (dash-dotted black line). For each altitude between 0 and 8 km asl, candlesticks show minimums, maximums and 1-sigma deviation from the mean of the density at each atmospheric layer. See Table~\ref{tab:sites} for a summary of the characteristics of each site. Altitudes were slightly shifted for the sake of clarity}\label{fig:atms}
    \end{center}
\end{figure*}

Once the EMF components are defined, the next step is to obtain the atmospheric profiles we shall use at each of the 23 sites.
For this calculation we use the monthly atmospheric profile for 2020 at each site, which was averaged from two local daily profiles extracted from the GDAS database and averaged following the ARTI methodology\,\cite{grisales2022impact}, obtaining $23\times12=276$ atmospheric profiles.
A sample of the obtained density profiles and their seasonal variations can be seen in the left panel of Figure~\ref{fig:atms}, where the seasonal density profiles of Los Alamos, are shown as a function of the altitude above sea level.
Density profiles follow the expected seasonal variations, with denser air at the ground level in winter and a decrease in the density in the summer's warm air.
In the right panel of the same Figure, expected variations along the year are shown for each atmospheric layer between ground level and $8$\,km asl.
These variations are characterised by the minimum, maximum and one sigma deviation from the mean observed during 2020.
We also included the variations observed at the High-performance Computing Center Stuttgart (HLRS, 453 m asl), the Centre de Calcul Recherche et Technology (CCRT, 94 m asl) and the Minho Advanced Computing Centre (MACC, 207 m asl) for comparative analysis.
The observed differences in the density profiles along the year are small, at the level of a few per cent, but they are critical when observing the atmospheric radiation at the ground level, as the atmospheric depth at a given altitude $h_i$, defined as the integral of the atmospheric density profile within the atmospheric layer of thickness $\delta h_i$, $X(h_i)=\int_{\delta h_i} \rho(h')\ \text{d}h'$, has a direct impact on the particles production, interactions and absorption at each particular layer (especially for altitudes below $\sim 15$\,km asl), and therefore, on the final secondary particle distribution at the ground.

Given the stochastic nature of the development of the EAS, a large sample of showers is needed to observe these effects on the expected flux at the ground in a statistically significant manner.
So the third step in the calculation is to integrate the primary spectrum $j$ to determine the total number of primary cosmic rays $N(A,Z) = \int j\ \text{d}\Omega\ \text{d}t\ \text{d}S\ \text{d}E_p$ of each relevant nucleus (identified by its atomic mass $A$ and number $Z$), which needs to be injected for a given integration time $t$, observation area $S$, solid angle interval $\Omega$, and primary energy $E_p$ range.

The cosmic ray energy spectrum ranges from GeV and up to more than $100$\,EeV and can be very well approximated by a simple monotonically decreasing power law, i.e.,
\begin{equation}
    \Phi(E_p, A, Z) \simeq \Phi_0(E_0, A, Z) \times (E_p/E_0)^{\alpha(E_p, A, Z)}, \label{eq:prim_flux}
\end{equation}
where $\Phi(E_p)$ is the expected flux of the considered primary nucleus $(A,Z)$, $\Phi_0$ is the reference flux at a certain energy $E_0$ for this particular nucleus, and $\alpha$ is the spectral index that depends on the primary energy and, while it can slightly vary from nucleus to nucleus, it can be well approximated by $\alpha \approx -3$ for the whole spectrum.
Thus, we can use this property of the primary flux to limit the upper energy limit when calculating the total number of primaries for each species that need to be injected.
Even more, at the PeV scale, the spectral index becomes steeper in the so-called {\emph{knee}} of the cosmic ray spectrum, i.e., $\alpha \approx -3.3$ at $E_p=4.5$\,PeV\,\cite{hoerandel2003knee}.
At the lowest energies, primaries are much more abundant but secondary particle production is limited and most of them are absorbed by the atmosphere before reaching ground level.
For all these reasons, we limit the primary energy range for the calculation of the expected background at the ground to $E_{\min} < E_p < 10^6$\,GeV, where $E_{\min}=m(Z,A) c^2+ 0.1$\,GeV, being $m(A,Z)$ the mass of the injected primary\,\cite{sarmiento2022arti}.

The second important parameter to be considered is the total integration time $t$.
While lower times reduce the total number of primaries needed to be simulated, the risk of the calculation being dominated by a statistical fluctuation increases as $t$ decreases.
So, in the end, a compromise has to be taken between the saving of computing resources and the statistical significance of the calculations.
While typical values for $t$ in astrophysics studies are up to a few hours\,\cite{asorey2018preliminary, sarmiento2019modeling}, in this case, we want to evaluate the atmospheric impact on the flux of secondary particles, and so we considered a total integration time $t$ of 1.5 days, i.e., $t=129,600$\,s for each month at $S=1$\,m$^2$ in each one of the $23$ sites to reduce statistical fluctuations.

Finally, since at these energies the primary flux is isotropic, we considered all the primaries following a uniform distribution in solid angle for the complete sky hemisphere around each site, i.e., $-\pi \le \varphi \le \pi$ and $0 \le \theta \le \pi/2$ for the local azimuth and zenith angle respectively.

Once the integration intervals are defined, the expected primary flux is integrated for all the relevant cosmic nuclei, obtaining $N\simeq 1.6\times 10^9$ primaries from protons to irons ($1\leq Z \leq 26$) for each month at each site, resulting in $\simeq 4.3\times 10^{11}$ simulated showers in $12\times23=276$ individual runs.
Calculations and analysis were done using the ARTI framework v1r9\,\cite{asorey2022arti}, including CORSIKA v7.7402\,\cite{heck1998corsika} for the EAS simulations, and QGSJET-II-04\,\cite{ostapchenko2011monte} and GEISHA-2002 libraries for accounting for the high- and low-energy interactions respectively.
The total flux of secondaries, $\Xi_{\text{All}}$, ranges from $\sim 700$ to $\sim 2,000$ particles per square metre per second, depending mainly on the EMF conditions, affecting the low energy sector of the primary flux\,\cite{asorey2018preliminary}; and the atmospheric profile, having a direct influence on particle production and absorption.
All the computations were performed on the ACME (equipped with Intel Gold 6138 processors) and TURGALIUM (Intel Gold 6254) clusters, demanding $\sim 450$\,kCPU$\cdot$hours and occupying a storage space of $1$\,TB for the final binary compressed files.

\begin{figure*}[!ht]
    \begin{center}
        \includegraphics[width=0.495\textwidth]{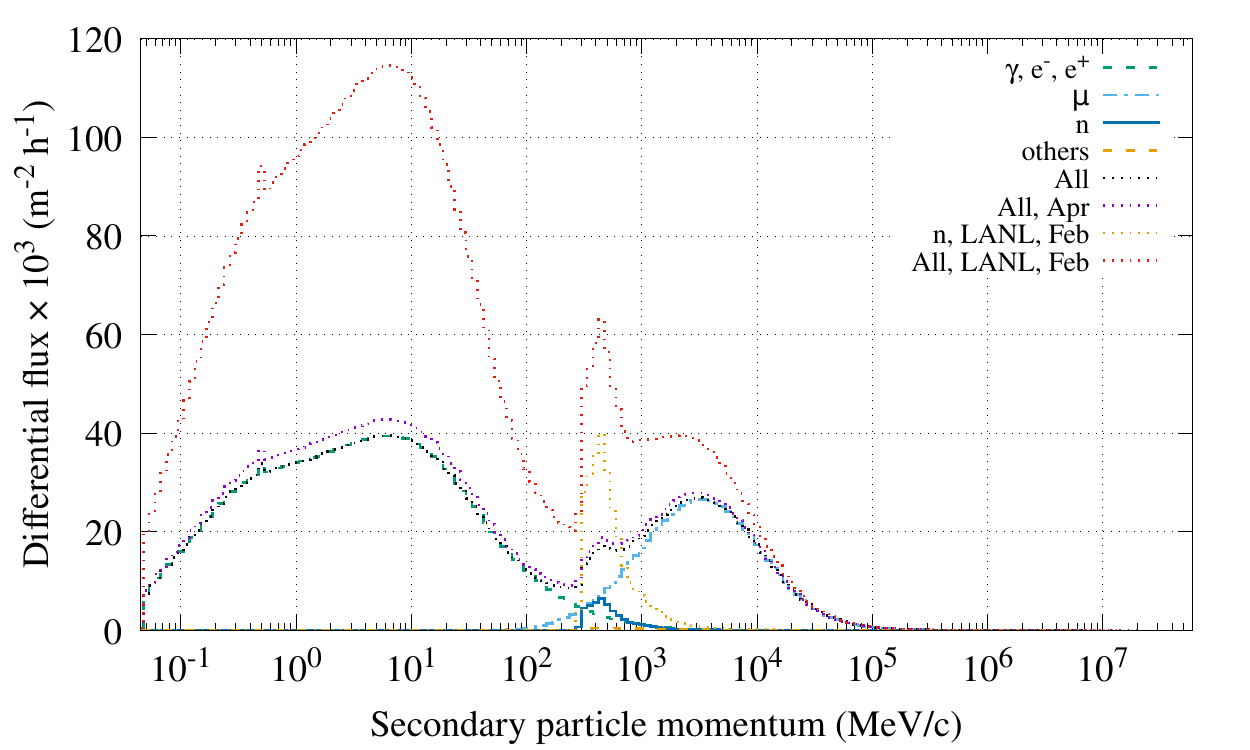}
        \includegraphics[width=0.495\textwidth]{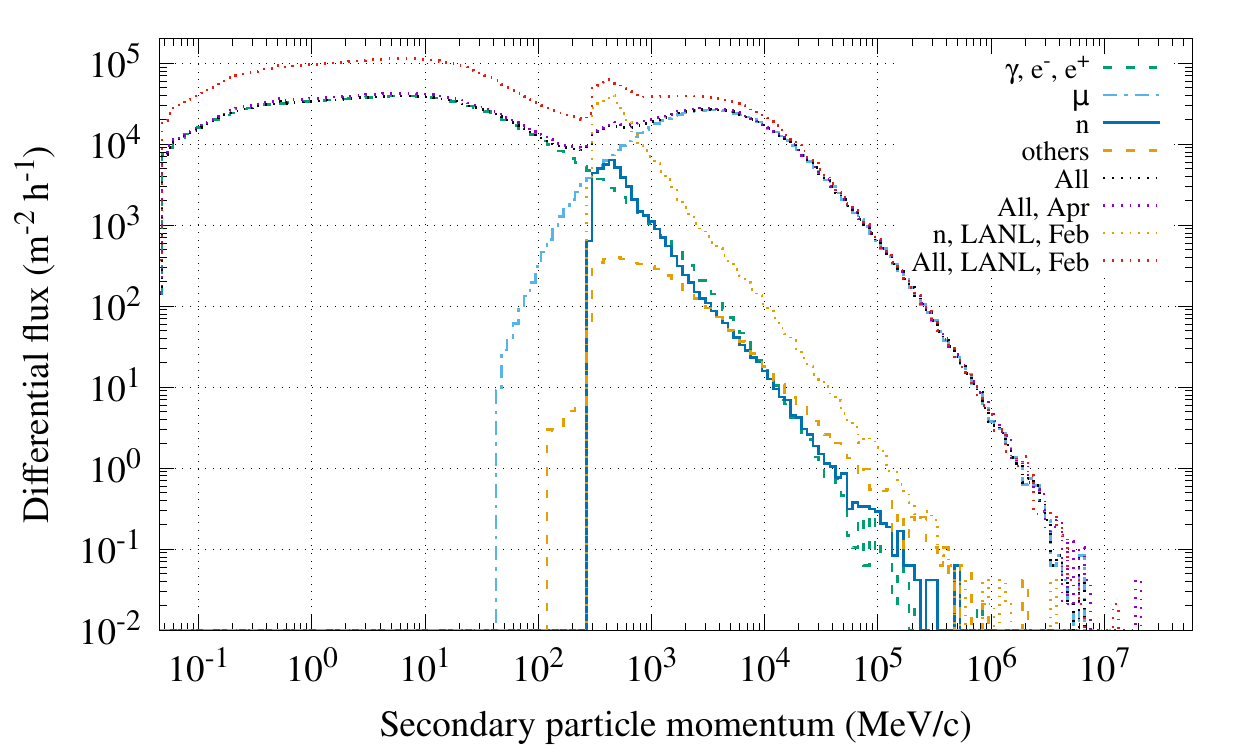}
        \caption{Linear-Log (left) and Log-Log (right) distribution of the momentum of the secondary particles $p_s$ expected in February 2020 at the ground level in MACC ($207$\,m asl). The main components of the showers, i.e., the electromagnetic component (dot-dashed green line), the muons $\mu^\pm$ (dot-long-dashed light blue line) and the neutrons (solid blue line) and other hadrons (dot-dashed yellow line), are identifiable by their own characteristics as described in the text. Total flux for February 2020 (dotted black line) and April 2020 are also show to evidence the seasonal effects. However, the major impact is produced by the altitude above sea level, as can be seen by comparison with the neutron (dotted yellow line) and total (dotted red line) fluxes expected in February 2020 at the ground level in LANL ($2,125$\,m asl)}\label{fig:histograms}
    \end{center}
\end{figure*}

Typically, secondary particles are grouped into three main groups: the electromagnetic component, composed of $\gamma$s and $e^\pm$, the hadronic component composed of neutrons, protons, nuclei and other baryons and mesons, and the muon $\mu^\pm$ component.
In Figure~\ref{fig:histograms}, the secondary momentum $p_s$ spectra are shown for these different components for the Minho Advanced Computing Centre (MACC) in Portugal, at an altitude of $200$\,m asl in February 2020.
Several important features of the cascade development can be inferred from this Figure.
At low $p_s$ values the flux is dominated by the electromagnetic (EM) component.
As explained in the previous section, as the shower evolves in the atmosphere, more and more energy is transferred to the EM component via particle decay and radiative processes.
However, EM particles are coupled to each other through different radiative processes and, thus, EM becomes the most important component of the shower development.
In the left panel of the Figure~\ref{fig:histograms}, a significant increase in the photon flux at the $510-520$\,keV energy bin is seen, corresponding to the production of $E_\gamma=511$\,keV photons via pair annihilation $e^+ e^- \to \gamma \gamma$ processes in the atmosphere.

The high-energy flux, shown in the right panel of Figure~\ref{fig:histograms}, is dominated by muons, charged leptons that carry the same interaction charges as $e^\pm$ but they are $\sim 200$ times as massive.
Thus, energy losses are relatively small compared with their typical energies: $\text{d}E/\text{d}X$ is in the range of $2-6$\,MeV\,cm$^2$\,g$^{-1}$, i.e.,$5-15$\,MeV\,cm in silicon, for muons in the $10^0-10^3$\,GeV energy range\,\cite{groom2001muon}.
Muons at the TeV scale, as those observed in Figure~\ref{fig:histograms}, possess enough energy to traverse hundreds and up to thousands of metres of rock and could be the main source for signals in muography studies\,\cite{bonechi2020atmospheric} or background noise at underground laboratories\,\cite{perez2022estimation}.
For the same reason, it is almost impossible to shield critical devices from muons, where they could induce SET and SEU soft errors by ionization for both types of muons, plus nuclear capture only for low-energy negative muons ($\sim 50\%$ of the total muon flux).
Recent works started to analyse the impact of atmospheric muons producing soft errors in different types of devices\,\cite{liao2019negative, hashimoto2020soft}.

Finally, at intermediate values of $p_s$, the non-thermal flux of atmospheric neutrons produces an important contribution to the total flux, especially at high-altitude sites.
The impact of the altitude and local atmospheric conditions can also be seen in the same Figure, where we also included the total flux of secondaries at the MACC site but for April 2020, and at Los Alamos National Laboratory (LANL, US, 2125 m asl) for February 2020.
Except for the flux of high-energy muons, which are essentially not affected by atmospheric absorption, the altitude effect is, by far, the dominant one when comparing the flux between different sites.
An increase of up to 3 times in the flux of secondary particles can be observed between the MACC and LANL sites.
It is also noticeable a lower but still statistically significant change in the flux originated from the change in the atmospheric profile at MACC between February and April 2020.

\begin{figure*}[!ht]
    \begin{center}
        \includegraphics[width=0.495\textwidth]{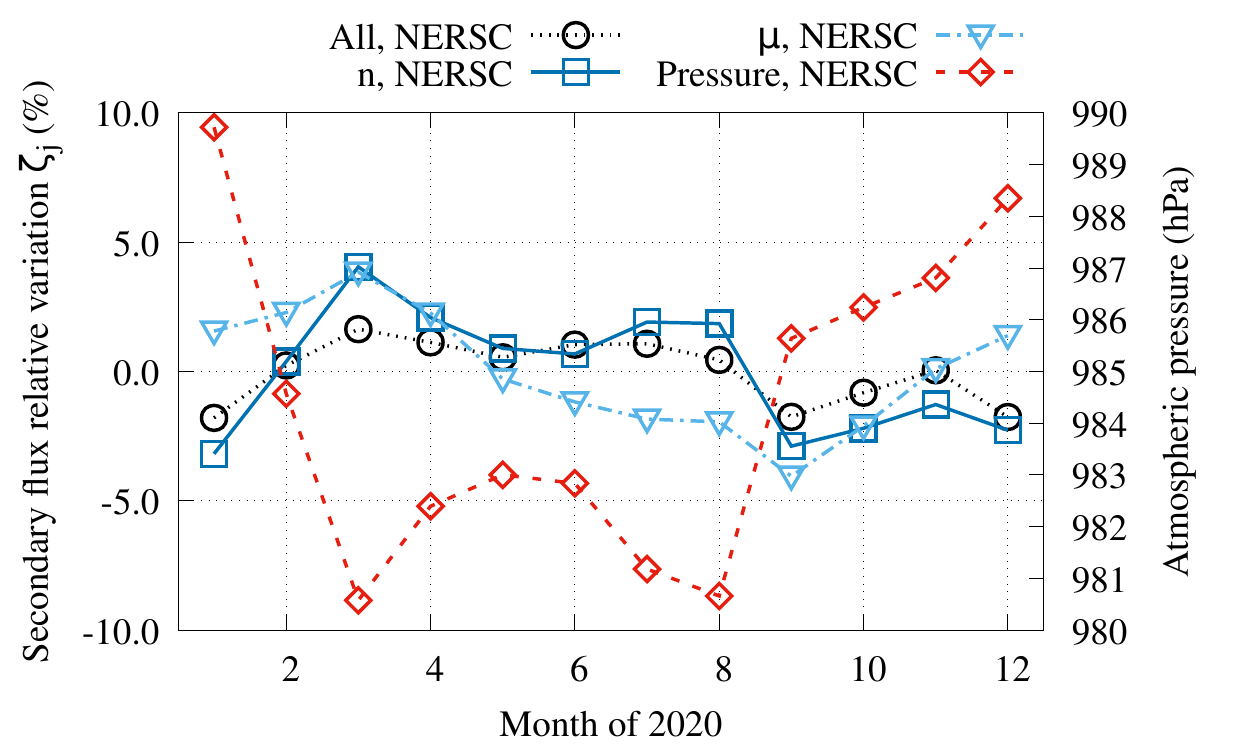}
        \includegraphics[width=0.495\textwidth]{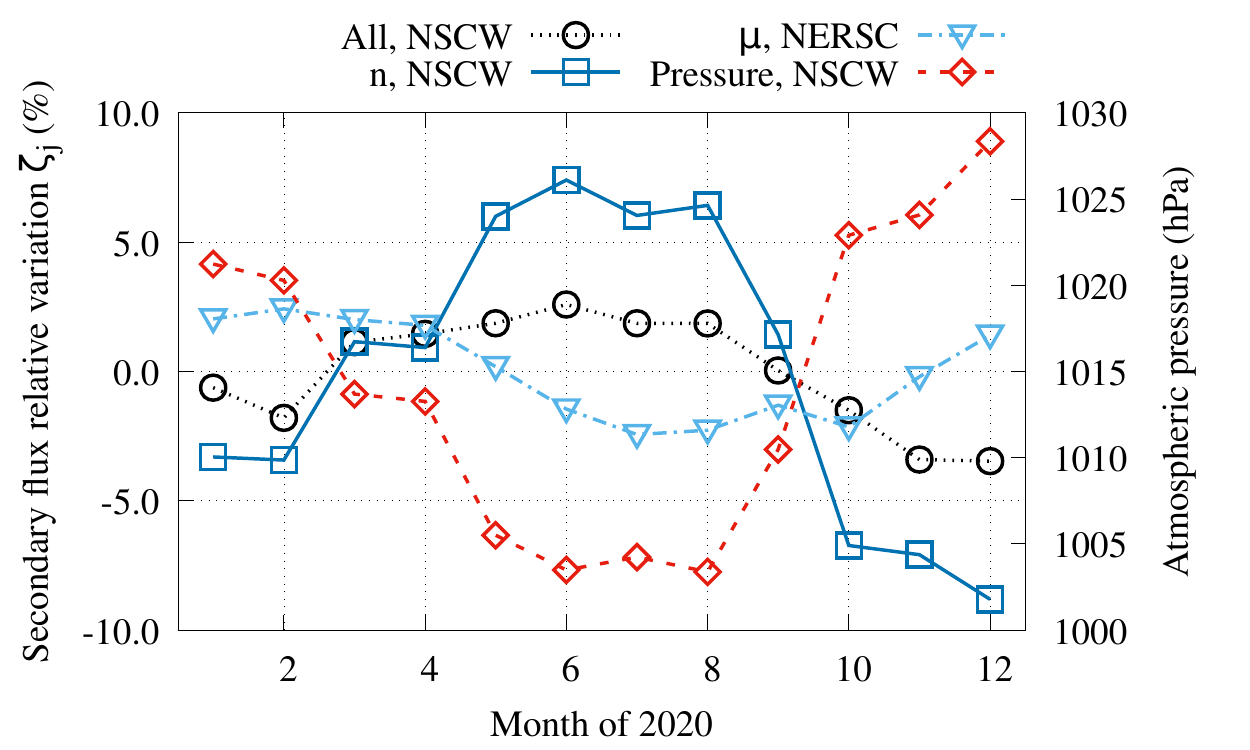}
        \caption{Expected relative flux variations $\zeta_j$ for neutrons (blue solid line, empty squares), muons (light blue dot-dashed line, empty triangles) and all the secondaries (black dotted line, empty circles); and the local atmospheric pressure at the ground (red dashed line, empty rhombus, right axis), are shown for each month of 2020 at the data centres of the National Energy Research Scientific Computing Center (NERSC, USA, $h_0\simeq 210$\,m asl) and National Supercomputing Center in Wuxi (NSCW, sea level, right). As described in the text, except for muons, the anti-correlation is remarkable at all the studied sites, especially for the neutron flux}\label{fig:anticorr}
    \end{center}
\end{figure*}

A denser atmosphere shall produce more absorption during the final stages of the development of the EAS, and so, a lower number of secondary particles at the ground will be observed, producing the well known anti-correlation between the atmospheric pressure and the rate of particles at the ground level\,\cite{dasso2012scaler}.
The atmospheric effect can be easily observed when studying the atmospheric pressure $P(h_0)$ at the ground level\footnote{Atmospheric pressure at a certain altitude $P(h)$ can be obtained from the atmospheric profiles by simply integrating the density profile, i.e., $P(h)=\int_{\infty}^{h} g \rho(h') \text{d}h'$, where $g$ is the acceleration due to gravity.} and the relative temporal variations in the expected flux of secondary type $j$, i.e.,
\begin{equation}
\zeta_j = \frac{\Delta \Xi_j}{\overbar{\Xi_j}} = \frac{\Xi_j(t)}{\overbar{\Xi_j}} - 1, \label{eq:zetaj}
\end{equation}
where $\Xi_j(t)$ is the instantaneous flux at time $t$ and $\overbar{\Xi_j}$ is the reference flux.
In Figure~\ref{fig:anticorr}, the values for $\zeta_j$ for high-energy neutrons, muons and total number of secondaries are shown together with the atmospheric pressure at the ground for the supercomputing centres of the National Energy Research Scientific Computing Center (NERSC, USA, $h_0\simeq 210$\,m asl) and the National Supercomputing Center in Wuxi (NSCW, China, $h_0\simeq 10$\,m asl).

Depending on the secondary type, the atmospheric dependence could be more or less important.
For example, in the right panel of Figure~\ref{fig:anticorr} the flux of electromagnetic particles is reduced due to the air absorption in the denser layers of the low atmosphere and, thus, the barometric modulation in Wuxi for the total flux is not as large as for neutrons.
Instead for muons, atmospheric absorption effect can be considered negligible, as can be appreciated in both panels of Figure~\ref{fig:anticorr}, where even a correlation can be observed during part of the year at some sites.
This can be explained by recalling that muons are mainly produced after charged pions decay, and so, local changes in density profiles at the muon production atmospheric depth are more relevant than the integral effect, that is related with the absorption.

\begin{figure*}[!ht]
    \begin{center}
        \includegraphics[width=0.495\textwidth]{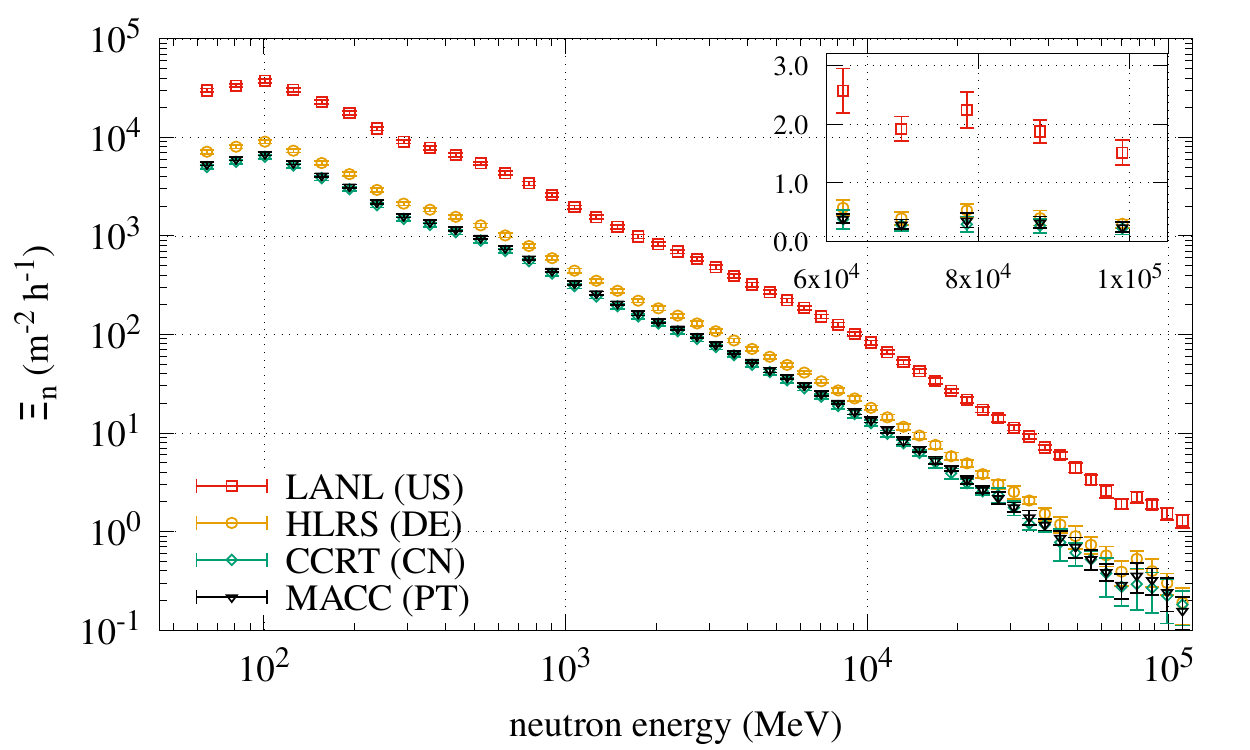}
        \includegraphics[width=0.495\textwidth]{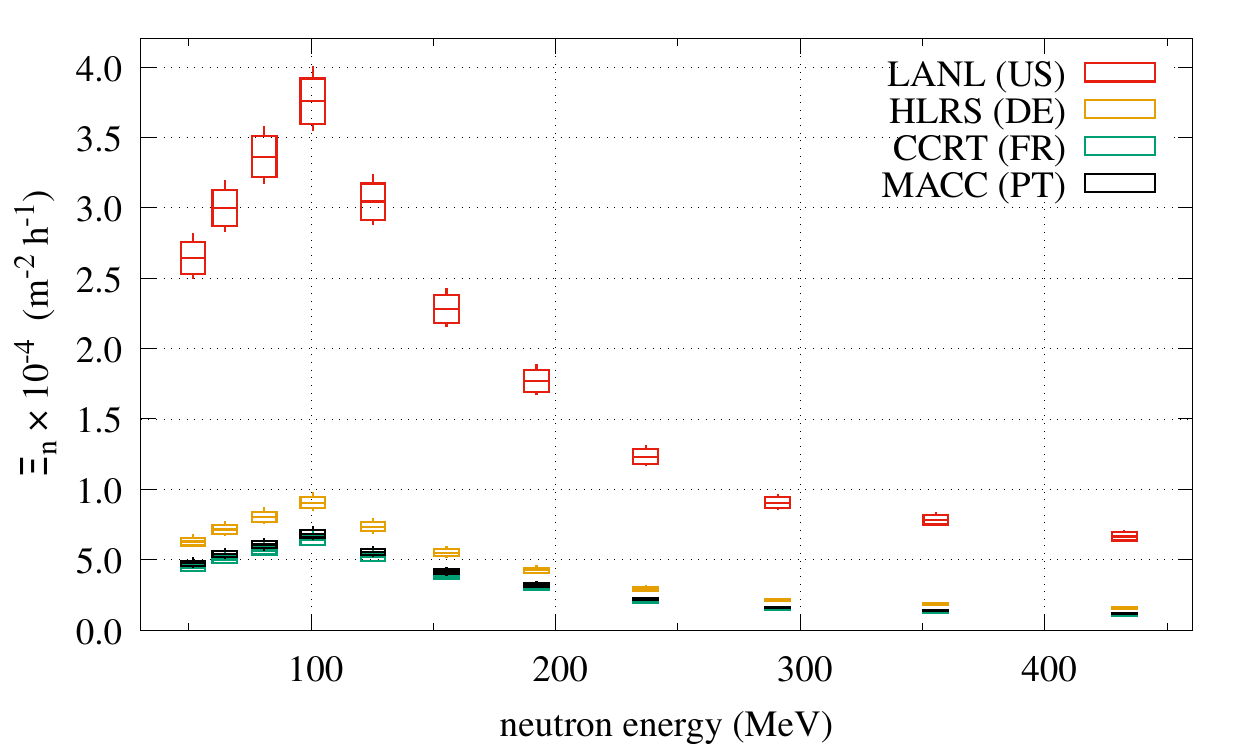}
        \caption{Energy distribution of the expected flux of neutrons $\Xi_n$ and its variations along 2020 at four sites: LANL (red squares), HLRS (yellow circles), CCRT (green rhombuses) and MACC (black triangles). At left, the $\Xi_n$ in the energy range $50<E_n<10^5$\,MeV is shown as long as the 1-sigma observed variation along the year. A slight increment in the flux is observed at $E_n\simeq 80$\,GeV (inset), consistent with neutron-nucleon cross-section increase at this energy range. The significant peak in $\Xi_n$, observed at $E_n\simeq 100$\,MeV, is detailed in the right panel, where the mean, $1\sigma$ deviations and the extrema in $\Xi_n$ for each energy bin are also shown. It can be noticed that the flux within each energy bin is not symmetric to the mean.}\label{fig:neu-histo}
    \end{center}
\end{figure*}

On the other hand, the atmosphere has a greater impact on neutron production, propagation, moderation and absorption, as can be also seen in Figure~\ref{fig:neu-histo}, where the average, deviation and extrema in the expected number of neutrons at the ground per squared metre and hour are shown as a function of their energy for the complete year of 2020 at four sites: Los Alamos National Laboratory (LANL, 2125 m asl), High-performance Computing Center Stuttgart (HLRS, 453 m asl), Centre de Calcul Recherche et Technology (CCRT, 94 m asl) and Minho Advanced Computing Centre (MACC, 207 m asl).
While the altitude effect is still dominant, the seasonal atmospheric variations have a noticeable effect on the flux of these high-energy neutrons ($E_n > 50$\,MeV), even at higher energies.
A detailed view of the $60-110$\,GeV neutron energy range is included, where a slightly significant deviation from the averaged power law is observed at $E_n\simeq 75)$\,GeV for all the sites.
This deviation is originated from the convolution of the decreasing energy at the production level with the increase in the neutron-nucleon cross-section at the $100$\,GeV scale\,\cite{ayre1975neutron}.

In the right panel of the same Figure, it is detailed the flux and its variations in the range $50\leqslant E_n/\text{MeV} < 450$, where the neutron flux increases by a factor of $1-2$ as the impact of the seasonal effects are enlarged.
At LANL, for example, the expected neutron flux in the $100$\,MeV energy bin could vary by $+15\%$, from $3.5\times 10^4$ up to $4.0\times 10^4$ neutrons per hour per squared metre, due only to the seasonal effect.

\begin{figure*}[!ht]
    \begin{center}
        \includegraphics[width=0.80\textwidth]{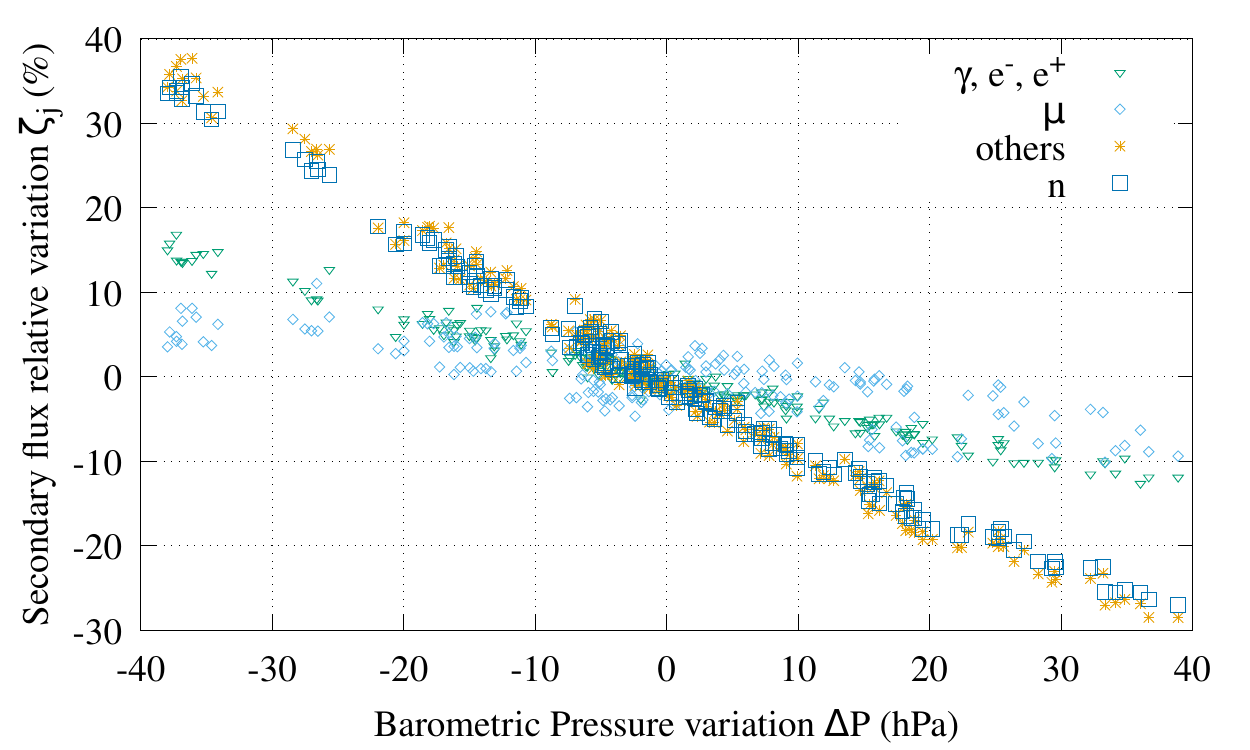}
        \caption{Effect of the changes in the barometric pressure in the expected flux of electromagnetic radiation (green triangles), muons (light blue rhombuses) and neutrons (blue squares) and other hadrons (yellow stars)   at the ground level for low altitude data centres ($h<1,000$\,m asl). Large variations are observed in the neutron flux for slightly small changes in the pressure. Due to their different atmospheric development, each type of particle evidence a very different response to changes in the barometric pressure, as it is evidenced in the slopes of these curves. For muons, on the other hand, local variations are most influenced by changes in the atmospheric profile at muon production layers than the barometric pressure. The exponential atmospheric dependence of the flux is visible in the slight deviation from a straight line.}\label{fig:pressure}
    \end{center}
\end{figure*}

To get a quantitative measure of the impact of the temporal variations of the atmosphere, in Figure~\ref{fig:pressure} the relative variation in the flux $\zeta_j$ for different types of secondaries $j$ is shown as a function of the variation of the local atmospheric pressure at all the low-altitude ($h<1,000$\,m asl) data centres.
The barometric effect has a different impact on each type of component of the showers due to their different development in the atmosphere.
This is visible in this Figure from the large differences in the observed slopes for each type of particle.
The biggest impact is for neutrons and other hadrons, evidencing global variations of up to $+40\%$ for a $-4\%$ decrease in the atmospheric pressure, with the flux $\Xi_n$ ranging from $42,500$ up to $68,500$ neutrons per squared metre per hour.

It is important to notice that, besides the obvious influence of the temperature on the air density, it also impacts the single shower distribution of particles at the ground due to local changes in the lateral development of the cascade\,\cite{abraham2009atmospheric}.
However, we are not interested in studying single EAS but looking for the global effect over the development of whole primary flux in the air producing the atmospheric radiation at the ground.
So, given the GCR flux isotropy and uniformity at the relevant energy ranges for this study, and the stochastic (Poissonian) and self-similarity\,\cite{sarmiento2022arti} nature of the atmospheric radiation production, the only effect that needs to be considered is related to the integral variation of the air density profile, i.e., the atmospheric pressure at the ground level.

Thereby, it is possible to take advantage of these effects to anticipate the expected flux of neutrons in different energy ranges at each data centre facility just by simply using the local atmospheric pressure at the ground as a tracer for the expected number of neutrons.

Local variations at each site are not as large as those shown in Figure~\ref{fig:pressure}, where all the observed seasonal variations with the global mean of the barometric pressure and the flux for each type of particle are shown together for the 22 low-altitude ($h<1,000$) data centres.
The slight deviation from the straight line is evidence of the exponential dependence of the flux of any secondary particle $j$, $\Xi_j(t)$, with the local barometric pressure $p(t)$ at time $t$.
Nevertheless, the observed variations in the barometric pressure at every single site are significantly smaller than the global ones, and thus, they can be modelled by:
\begin{equation}
    \zeta_j = \beta_j \Delta P,
    \label{eq:betaj}
\end{equation}
where $\beta^{j}$ is the barometric coefficient for secondary $j$ and $\Delta P = P(t) - \overbar{P}$ is the variation of the atmospheric pressure to the local reference $\overbar{P}$.
As this can be also done for different energy ranges, in this work we considered three different ones: the complete simulated energy range, $E_n\geqslant50)$\,MeV; $(50 \leqslant E_n \leqslant 1,000)$\,MeV; and $(E_n > 1,000)$\,MeV; respectively labelled as $i=0$, $i=1$ and $i=2$.
For the sake of clarity and given we are mainly focused on the neutron flux, we can obviate the subscript $n$ and so, equation~\eqref{eq:betaj} could be written as:
\begin{equation}
    \zeta_i = \beta_i \Delta P,
    \label{eq:beta}
\end{equation}
where now the subscript $i$ refers to the corresponding neutron energy range $i=0,1,2$ described above.
The obtained results for all the sites are compiled in the table~\ref{tab:neutrons}.
It is important to notice that slight differences could be observed in both the total flux and the barometric coefficients at sites with similar altitudes due to differences of the atmospheric profiles and their impacts on the neutron flux.

\begin{center}
    \begin{sidewaystable}
        \centering
        \caption{Altitude, latitude and longitude of the 23 new exascale facilities, including the total, averaged flux $\overbar{\Xi}$ in m$^{-2}$\,hour$^{-1}$ of all the secondary particles, muons and neutrons with $E_n\geqslant 50$\,MeV expected at the ground level in each site.}\label{tab:sites}
        \scriptsize
        \begin{tabular}{p{5.4cm}p{1.7cm}rrrlcccc}
            \toprule
            {\textbf{Supercomputing centre}} & Country & Alt. & Lat. & Lon. & Code & $\Xi_{All} \times 10^{-6}$ & $\Xi_{n} \times 10^{-4}$ & $\Xi_{\mu} \times 10^{-5}$ \\
            \midrule
            Los Alamos National Laboratory &USA &2,125 &35.85 &-106.29 &LANL &$(6.70 \pm 0.19)$ &$(26.4 \pm 1.1)$ &$(9.7 \pm 0.3)$ \\
            National University of Defense Technology &China &750 &27.93 &107.71 &NUDT &$(3.43 \pm 0.08)$ &$(7.7 \pm 0.4)$ &$(7.4 \pm 0.1)$ \\
            Centro de Investigaciones Energéticas, Medioambientales y Tecnológicas &Spain &700 &40.50 &-3.67 &MAD &$(3.47 \pm 0.06)$ &$(7.7 \pm 0.3)$ &$(7.7 \pm 0.1)$ \\
            Sofia Tech Park &Bulgaria &565 &42.67 &23.37 &SOFIA &$(3.32 \pm 0.05)$ &$(6.9 \pm 0.2)$ &$(7.6 \pm 0.1)$ \\
            Leibniz Supercomputing Centre &Germany &471 &48.26 &11.67 &LRZ &$(3.21 \pm 0.07)$ &$(6.4 \pm 0.3)$ &$(7.5 \pm 0.1)$ \\
            High-performance Computing Center Stuttgart &Germany &453 &48.74 &9.10 &HLRS &$(3.19 \pm 0.07)$ &$(6.3 \pm 0.3)$ &$(7.5 \pm 0.1)$ \\
            Institute of Information Science &Slovenia &280 &46.56 &15.65 &IZUM &$(2.97 \pm 0.05)$ &$(5.3 \pm 0.2)$ &$(7.3 \pm 0.1)$ \\
            uxConnec's Data Center DC2 &Luxembourg &275 &49.79 &6.09 &DC2 &$(2.98 \pm 0.07)$ &$(5.3 \pm 0.2)$ &$(7.3 \pm 0.1)$ \\
            IT4 Innovations National Supercomputing Center &Czechia &261 &49.84 &18.16 &IT4 &$(2.96 \pm 0.07)$ &$(5.2 \pm 0.2)$ &$(7.3 \pm 0.1)$ \\
            Oak Ridge National Laboratory &USA &250 &35.93 &-84.31 &ORNL &$(2.85 \pm 0.05)$ &$(4.8 \pm 0.1)$ &$(7.0 \pm 0.1)$ \\
            Argonne National Laboratory &USA &214 &41.72 &-87.98 &ANL &$(2.87 \pm 0.04)$ &$(4.9 \pm 0.2)$ &$(7.1 \pm 0.2)$ \\
            National Energy Research Scientific Computing Center &USA &210 &37.88 &-122.25 &NERSC &$(2.85 \pm 0.03)$ &$(4.8 \pm 0.1)$ &$(7.0 \pm 0.2)$ \\
            Minho Advanced Computing Centre &Portugal &207 &41.56 &-8.40 &MACC &$(2.85 \pm 0.05)$ &$(4.8 \pm 0.2)$ &$(7.1 \pm 0.1)$ \\
            Lawrence Livermore National Laboratory &USA &188 &37.69 &-121.70 &LLNL &$(2.84 \pm 0.03)$ &$(4.7 \pm 0.1)$ &$(7.0 \pm 0.1)$ \\
            Datacenter CSC Kajaani &Finland &128 &64.23 &27.70 &CSCF &$(2.94 \pm 0.08)$ &$(5.1 \pm 0.3)$ &$(7.4 \pm 0.2)$ \\
            Barcelona Supercomputing Center &Spain &100 &41.39 &2.12 &BSC &$(2.76 \pm 0.05)$ &$(4.3 \pm 0.2)$ &$(7.0 \pm 0.1)$ \\
            Jülich Supercomputing Centre &Germany &100 &50.92 &6.54 &JSC &$(2.79 \pm 0.05)$ &$(4.5 \pm 0.2)$ &$(7.1 \pm 0.1)$ \\
            Poznan Supercomputing and Networking Center &Poland &100 &52.41 &16.92 &PSNC &$(2.80 \pm 0.05)$ &$(4.5 \pm 0.2)$ &$(7.1 \pm 0.1)$ \\
            Centre de Calcul Recherche et Technologie &France &94 &48.60 &2.20 &CCRT &$(2.78 \pm 0.05)$ &$(4.4 \pm 0.2)$ &$(7.1 \pm 0.1)$ \\
            Bologna Technopole &Italy &40 &44.52 &11.36 &BOLT &$(2.72 \pm 0.06)$ &$(4.2 \pm 0.2)$ &$(7.0 \pm 0.1)$ \\
            National Supercomputer Center in Guangzhou &China &10 &23.07 &113.39 &NSCG &$(2.58 \pm 0.05)$ &$(3.7 \pm 0.2)$ &$(6.4 \pm 0.1)$ \\
            National Supercomputing Center in Wuxi &China &10 &31.57 &120.30 &NSCW &$(2.60 \pm 0.05)$ &$(3.8 \pm 0.2)$ &$(6.6 \pm 0.1)$ \\
            RIKEN Center for Computational Science &Japan &10 &34.68 &135.22 &RCCS &$(2.63 \pm 0.04)$ &$(3.9 \pm 0.1)$ &$(6.7 \pm 0.2)$ \\
            \bottomrule
        \end{tabular}
    \end{sidewaystable}
\end{center}

\begin{table}[ht]
    \caption{Reference pressure $\overbar{P}$ and neutron flux $\overbar{\Xi_i}$, and barometric coefficients $\beta_i$ (in hPa$^{-1}$) for the $i$-th energy range at the 23 exascale facilities. With these values, it is possible to calculate using equation~\eqref{eq:beta} the local variations in the flux of neutrons just from the local barometric pressure (use $\overbar{\Xi_0}$ and $\beta_0$ for the neutron flux with $E_n\geqslant 50$\,MeV). Pressure is given in hPa and fluxes are given in m$^{-2}$\,hour$^{-1}$.}\label{tab:neutrons}
    \scriptsize
    \begin{center}
        \begin{tabular}{lrrrrrrrr}
            \toprule
            {\textbf{Site}} & Alt. & {$\overbar{P}$} & \textbf{$\overbar{\Xi_0}$} & {\textbf{$\beta_0$}} & {$\overbar{\Xi_1}$} & {\textbf{$\beta_1$}} & {$\overbar{\Xi_2}$} & {\textbf{$\beta_2$}} \\
            \midrule
            LANL &2,125 &777 &26.4 &-9.2 &25.7 &-9.2 &7.0 &-9.7 \\
            NUDT &750 &927 &7.7 &-6.9 &7.5 &-6.9 &1.9 &-7.1 \\
            MAD &700 &927 &7.7 &-7.7 &7.5 &-7.7 &1.9 &-7.5 \\
            SOFIA &565 &940 &6.9 &-6.8 &6.8 &-6.8 &1.7 &-7.0 \\
            LRZ &471 &950 &6.4 &-7.8 &6.2 &-7.8 &1.6 &-7.9 \\
            HLRS &453 &952 &6.3 &-8.1 &6.1 &-8.1 &1.6 &-8.3 \\
            IZUM &280 &974 &5.3 &-6.8 &5.2 &-6.8 &1.3 &-7.0 \\
            DC2 &275 &973 &5.3 &-7.9 &5.2 &-7.9 &1.3 &-8.2 \\
            IT4 &261 &975 &5.2 &-7.6 &5.1 &-7.6 &1.3 &-7.8 \\
            ORNL &250 &984 &4.8 &-7.9 &4.7 &-7.9 &1.2 &-8.2 \\
            ANL &214 &983 &4.9 &-8.8 &4.8 &-8.8 &1.2 &-9.6 \\
            NERSC &210 &984 &4.8 &-7.0 &4.7 &-7.0 &1.2 &-6.4 \\
            MACC &207 &986 &4.8 &-7.6 &4.6 &-7.6 &1.2 &-8.0 \\
            LLNL &188 &987 &4.7 &-7.1 &4.6 &-7.1 &1.1 &-7.7 \\
            CSCF &128 &978 &5.1 &-8.4 &5.0 &-8.4 &1.3 &-8.7 \\
            BSC &100 &997 &4.3 &-7.7 &4.2 &-7.7 &1.1 &-7.9 \\
            JSC &100 &993 &4.5 &-7.7 &4.4 &-7.8 &1.1 &-7.7 \\
            PSNC &100 &993 &4.5 &-7.3 &4.4 &-7.3 &1.1 &-7.7 \\
            CCRT &94 &995 &4.4 &-8.2 &4.3 &-8.2 &1.1 &-8.4 \\
            BOLT &40 &1002 &4.2 &-7.2 &4.1 &-7.2 &1.0 &-7.2 \\
            NSCG &10 &1015 &3.7 &-6.9 &3.6 &-6.9 &0.9 &-7.2 \\
            NSCW &10 &1014 &3.8 &-6.4 &3.7 &-6.4 &0.9 &-6.5 \\
            RCCS &10 &1010 &3.9 &-6.7 &3.8 &-6.7 &0.9 &-6.7 \\
            &      &    & $\times 10^4$ & $\times 10^{-3}$ & $\times 10^4$ & $\times 10^{-3}$ & $\times 10^4$ & $\times 10^{-3}$ \\
            \bottomrule
        \end{tabular}
    \end{center}
\end{table}

From these values and using~\eqref{eq:beta}, it is possible to estimate the expected flux of high-energy neutrons and its variations at each site just by measuring the local atmospheric pressure, since $\beta_i$ corresponds to the relative decrease (increase) in the neutron flux for an $1$\,hPa increase (decrease) in the local barometric pressure.
For example, from the second and the fourth column of Table~\ref{tab:neutrons}, the reference atmospheric pressure and the global barometric coefficient for the site of Los Alamos (LANL) are $\overbar{P}=777$\,hPa and $\beta_0=-9.2\times10^{-3}$\,hPa$^{-1}$ respectively.
Therefore, on a typical sunny day at LANL, when the barometric pressure should be higher than usual, say, $P(t)=779$\,hPa, a reduction of $\beta_0 (P(t)-\overbar{P}) = - 9.2\times10^{-3}\text{\ hPa}^{-1} \times 2\text{\ hPa} = -1.84\times10^{-2} \simeq -2\%$ in the $E_n \gtrsim 50$\,MeV neutron flux shall be expected.
Thunderstorms, on the other hand, are preceded by a drop in the atmospheric pressure of several hPa in a few hours, with typical drop rates of at least $-1$\,hPa\,h$^{-1}$.
So, at sea level, the barometric pressure could be as low as $1,002$\,hPa, or even less, during a thunderstorm.
Thus, for example, during the preclude of a thunderstorm at the RIKEN Center for Computational Science (RCCS) in Kobe, Japan, where the average atmospheric pressure is $\overbar{P}=1,010$\,hPa, an increase of $\sim 6\%$ in the flux of neutrons with energies above $50$\,MeV could be expected\footnote{Since, according to Table~\ref{tab:neutrons} for RCSS: $\beta_0 (P(t)-\overbar{P}) = -6.7\times10^{-3}\text{\ hPa}^{-1} \times (-8)\text{\ hPa} \simeq 6\%$.}, and the situation could be even worst when considering the effective moderation of neutrons produced by rain.
As a consequence, an increase (decrease) in the flux of high-energy neutrons will result in a similar increase (decrease) in the probability of errors produced in the supercomputer.

For muons, local changes in the profile at muon production depth are the dominant effect.
Expected average muon flux at each data centre for $E_\mu \geqslant 15$\,MeV are also included in Table~\ref{tab:sites}.

Space weather phenomena, such as the disturbances of the magnetosphere produced by the passage of an interplanetary coronal mass ejection (iCME) by Earth\,\cite{masias2016superposed}, also impacts the flux of high-energy neutrons and for this reason, atmospheric neutrons have been used since decades ago to monitor Solar activity\,\cite{mishev2020current}.
These phenomena are observed as decreases in the total flux of atmospheric neutrons, where reductions of up to $35\%$ could be expected for $E_n\simeq 100$\,MeV neutrons during severe geomagnetic storms\,\cite{asorey2018preliminary}, and some astroparticle observatories, such as LAGO, are focused on enhancing their neutron detection capabilities\,\cite{sidelnik2020enhancing,sidelnik2020simulation}.

These scenarios are important when anticipating possible errors associated with the flux of high-energy neutrons at supercomputer centres, as it will be discussed in subsection~\ref{subsec:fit}.

\subsection{High-energy neutrons modulations and soft error rates at supercomputers}\label{subsec:fit}

A typical magnitude used to describe the device performance in terms of its sensitivity to radiation is the FIT (failures-in-time) rate, i.e., the number of observed failures of a certain (or any) kind in $10^9$ (one billion) hours of device operation, and so, the total FIT is just the sum of each kind of failure: $\text{FIT}=\sum_k^N \text{FIT}_k$.
From this definition, the MTBF measured in hours is just the reciprocal of FIT times $10^9$:
\begin{equation}
    \text{MTBF} = \frac{10^9}{\text{FIT}}. \label{eq:MTBF}
\end{equation}

It is possible to obtain the FIT rate from the effective cross-section $\sigma_\text{err}$, as it is just an effective measure of the probability that a neutron triggers a certain type of error in a device, and it is typically expressed in units of area (cm$^2$)\,\cite{baumann2005radiation}.
Thus, in general,
\begin{equation}
\text{FIT}_{\text{err}}=10^{5}\ \Xi\ \sigma_{\text{err}}, \label{eq:fit}
\end{equation}
when the flux $\Xi$ is expressed in units of m$^{-2}$\,h$^{-1}$.
Then, by combining this result with equations~\eqref{eq:zetaj} and~\eqref{eq:beta} for neutrons:
\begin{equation}
\text{FIT}_{\text{err}}(t) = 10^{5}\ \sigma_{\text{err}}\ \overbar{\Xi_i} \left[ 1 + \beta_i \left(P(t) - \overbar{P} \right) \right], \label{eq:absfit}
\end{equation}
in the $i$-th neutron energy range, for pressure expressed in hPa and $\sigma$ in cm$^2$.

Oliveira {\textit{et al.}}\,\cite{oliveira2021thermal, oliveira2020high} irradiate different types of commercial off-the-shelf (COTS) devices by exposing them to neutron beams in energy scales from thermal to $1$\,GeV, obtaining the device sensitivity to neutrons measured through the identification of unrecoverable errors (DUE) or SDC in APUs (CPUs+GPUs integrated in the same device), FPGAs and DDR memories.
Unfortunately, they only present cross-sections \lq\lq{}relative to the lowest one measure for each vendor to prevent the leakage of business-sensitive data\rq\rq{}\,\cite{oliveira2021thermal}.
However, is it possible to see that, for all the tested devices, thermal neutron cross-sections are far for being negligible\cite{oliveira2021thermal}, but in most cases they are still considerable smaller than the corresponding effective cross-section of high-energy neutrons (the observed differences are up to one order of magnitude for APUs).
Similar conclusions can be obtained from Figure~6 of \,\cite{oliveira2020high}, where it is possible to observe that, in presence of the nominal atmospheric flux of high-energy neutrons ($E_n>10$\,MeV), the FIT rates are totally dominated by them.

As mentioned in section~\ref{sec:related-work}, Tiwari {\textit{et al.}}\,\cite{tiwari2015understanding}, analyzed the error logs of two GPU supercomputing facilities: the Titan supercomputer at the Oak Ridge National Laboratory (ORNL), consisting of $18,688$ K20X GPUs;
and of the Moonlight GPGPU cluster at Los Alamos National Laboratory (LANL), consisting of $616$ M2090 GPGPUs.
By exposing K20X GPU to the ISIS and LANSCE white neutron sources, that emulate the atmospheric neutron flux in the $10<E_n<750$\,MeV energy range\,\cite{violante2007new}, they were able to obtain the SDC and program crashes effective cross-sections $\sigma_{\text{err}}$, that are compiled in the table~2 of\,\cite{tiwari2015understanding} and, for the worst case scenario, they can be averaged obtaining $\sigma_{\text{SDC}} = (4.8 \pm 0.4) \times 10^{-7}$\,cm$^2$ and $\sigma_{\text{crash}} = (2.7 \pm 0.2) \times 10^{-7}$\,cm$^2$ respectively.
While the energy ranges of the neutron sources used for the irradiation of the K20s devices are lower than the complete energy range simulated in this work, it is possible to assume that the neutron-error cross-sections in the energy range $E_n>1,000$\,MeV should not be far from the reported values.
Moreover, at these high energies, the flux is considerably lower than in the $50 \leq E_n/\text{MeV} \leq 1,000$ energy range, and so the error rates will be dominated by the flux within this range.
Therefore, following equation~\eqref{eq:absfit} and using the tabulated values for $\overbar{P}$, $\overbar{\Xi_1}$ and $\beta_1$ for the ORNL site, the expected FIT$_{\text{SDC}}$ rate when the atmospheric pressure drops by, say, $-5$\,hPa respect to the barometric reference pressure, should be\footnote{$FIT_{\text{SDC}}=(10^5)(4.7\times10^4)(4.8\times10^{-7})[1+(-7.9\times10^{-3})(979-984)]=2,345\simeq 2,300$ failures in 10$^9$ device$\cdot$hours of operation.} of FIT$_{\text{SDC}} \sim 2,300$, and so, from equation~\eqref{eq:MTBF}, the corresponding MTBF for the whole Titan supercomputer should be of $\simeq 23$\,hours, i.e., about $1$ silent error per day due to the expected flux of neutrons with $50<E_n<1,000$\,MeV when the atmospheric pressure drops by $-5$\,hPa.

Once the expected flux of neutrons was determined for each site, calculation of effective flux at computing devices, including CPUs, GPUs, APUs, storage and memories, have to take into account the geometry and materials of computing racks, buildings and other infrastructures in the surroundings, even, on the supercomputing cooling system, especially those using water or any other aqueous solutions as coolants.
All these components will have a profound impact in the flux of high-energy neutrons, producing thermal and epi-thermal neutrons having different cross-sections with the materials used for making the different types of devices available in any data centre\@.

As a final remark, given the linearity of equations~\eqref{eq:beta} and~\eqref{eq:fit}, it is easy to see that the relative variation of the FIT rates,
\begin{equation*}
\psi_{\text{err}} = \frac{\text{FIT}_{\text{err}}(t)}{\overbar{\text{FIT}_{\text{err}}}}-1,
\end{equation*}
where $\overbar{\text{FIT}_{\text{err}}}$ is the reference FIT$_{\text{err}}$ rate at the site, is equal to the relative variation $\zeta$ of the high-energy neutron flux, i.e, $\psi=\zeta$, and so:
\begin{equation}
\psi_{\text{err}} = \beta \Delta P, \label{eq:fitp}
\end{equation}
that is, the FIT rate associated with the flux of high-energy neutrons at each site should evidence a small anti-correlation ($\beta > -1\%$) with the local changes in the barometric pressure that increases with the altitude of the supercomputing centre.

\section{Conclusions}\label{sec:conclusions}

In this work we presented the calculation of the expected flux of atmospheric neutrons of $E_n \geq 50$\,MeV and their seasonal variations at each one of the 23 future sites of the next generation of exascale supercomputing facilities.
This was done by simulating the interaction of the measured galactic cosmic rays flux and including real atmospheric and geomagnetic conditions at each site using the state-of-the-art techniques and codes heavily used, tested and validated in the astroparticle physics community.

By using real atmospheric profiles, extracted from the GDAS database and averaged to obtain the atmosphere conditions for each month of 2020, the expected flux of high-energy neutrons with $E_n \geq 50$\,MeV and its seasonal variations at each exascale supercomputing centre were obtained and parametrised.
The dependence on the total flux of particles and neutron flux with the atmospheric pressure was observed and the barometric pressure coefficient for neutrons at different energy ranges were obtained and they are summarised in Table~\ref{tab:neutrons}.
The reported barometric coefficients, $\beta_i$ corresponds to the relative change in the expected flux in different energy ranges when the atmospheric pressure changes by $\pm 1$\,hPa.
The provided information makes it possible to easily estimate the expected flux of neutrons under different atmospheric conditions (equation~\eqref{eq:beta}) and to evaluate the corresponding FIT rates of silent errors due to high-energy neutrons (equation~\eqref{eq:absfit}) and its relative seasonal variations (equation~\eqref{eq:fitp}).
This can be done by using the instantaneous barometric pressure that can be easily measured at each facility, being a simple and direct way to anticipate potential silent and non-silent errors that could appear during critical calculations that could be performed soon at the next generation of exascale supercomputing facilities.

To avoid the intrinsic limitation of CORSIKA for low energy neutrons, we are currently developing a special module in ARTI, based on FLUKA\,\cite{bohlen2014fluka}, to extend current calculations down to the meV neutron energy scale.
Extensions of the atmospheric flux simulations using real atmospheres presented here but including other effects such as the rain, that could double the thermal neutron flux at the ground as water droplets acts as neutrons moderators, and the corresponding Geant4\,\cite{agostinelli2003geant4} simulations of neutron moderation in infrastructures are being considered and will be published as a follow-up of the analysis presented here.

\section*{Declarations}

\subsection*{Availability of data and materials}

The datasets generated and analysed during the current study are available in the Zenodo repository, \href{https://doi.org/10.5281/zenodo.6721615}{DOI:10.5281/zenodo.6721615}.
The ARTI code is available in the LAGO GitHub repository: \href{https://github.com/lagoproject/arti}{github.com/lagoproject/arti}, \href{https://doi.org/10.5281/zenodo.7316555
}{DOI:10.5281/zenodo.7316555}.

\subsection*{Acknowledgments}

This work has been partially funded by the co-funded Spanish Ministry of Science and Innovation project CODEC-OSE (RTI2018-096006-B-I00) with European Regional Development Fund (ERDF) funds, by the co-funded European Union Horizon 2020 research and innovation Programme project EOSC-SYNERGY (grant agreement No 857647), and by the co-funded Comunidad de Madrid project CABAHLA-CM (S2018/TCS-4423).
Also, this work was partially supported by the computing facilities (Turgalium) of Extremadura Research Centre for Advanced Technologies (CETA-CIEMAT), funded by the ERDF too.

The authors are grateful to Antonio Juan Rubio-Montero and Angelines Alberto-Morillas from CIEMAT, Alfonso Pardo-Diaz from CETA/CIEMAT and Iván Sidelnik from CNEA for their continuous support and fruitful discussions.

HA thanks Rafael Mayo-García for his warm welcome and continuous support during his stay at CIEMAT in Madrid, Spain.

\bibliographystyle{ieeetr}
\bibliography{neutrons-cpc}
\end{document}